\documentclass[aps,prl,floatfix,reprint,bibliography]{revtex4-1}

\usepackage{amsmath}
\usepackage{amssymb}
\usepackage{graphicx}
\usepackage{placeins}
\usepackage{epstopdf} 
\usepackage{xcolor}
\usepackage{comment}

\newcommand{\etal}{\textit{et al. }}

\begin{document}


\title{Origin of gap-like behaviors in {U}{R}u$_2${S}i$_2$: Combined study via quasiparticle scattering spectroscopy and resistivity measurements}


\author{S. Zhang$^{1,2}$, G. Chappell$^{1,2}$, N. Pouse$^3$, R. E. Baumbach$^{1,2}$, M. B. Maple$^3$, L. H. Greene$^{1,2}$, W. K. Park$^{1,}$}
 \email{wkpark@magnet.fsu.edu}
\affiliation{$^1$National High Magnetic Field Laboratory, Florida State University, Tallahassee, Florida 32310\\
$^2$Department of Physics, Florida State University, Tallahassee, Florida 32306\\
$^3$Department of Physics, University of California, San Diego, California 92093}


\date{\today}

\begin{abstract}
We address two long-standing questions regarding the hidden order in URu$_2$Si$_2$: Is it associated with the hybridization process, and what are the distinct roles played by the localized and itinerant electrons? Our quasiparticle scattering spectroscopy reveals a hybridization gap ubiquitous in the entire phase space spanned by P and Fe substitutions in URu$_2$Si$_2$, including the no-order and antiferromagnetic regions, with minimal change upon crossing the phase boundary. This indicates its opening isn't associated with the ordering, and thus localized electrons must be the major player. Towards a consistent understanding of all the other gap-like behaviors observed only below transition temperatures, we analyze the electrical resistivity using a model in which gapped bosonic excitations are the dominant scattering source. With their stiffness set to follow an unusual temperature dependence (decreasing with decreasing temperature), this model fits all of our resistivity data well including the jump at the transition. Remarkably, the extracted gap increases slowly with increasing Fe content, similarly to the gap detected by inelastic neutron scattering at Q$_1$ = (1.4, 0, 0), suggesting a common origin. Such a model can also naturally explain the Hall effect temperature dependence without invoking Fermi surface gapping.
\end{abstract}
%


\maketitle

Strongly correlated electron systems oftentimes exhibit seemingly similar phase diagrams. For their comprehensive understanding, it is not only necessary to identify the underlying interactions but also to elucidate the interplay among them. The $f$-orbital based heavy-fermion compounds are an archetypal correlated system, in which the hybridization between itinerant and localized electrons causes the emergence of heavy fermions \cite{hewson1997kondo,coleman2007heavy}. What different roles are played by multiple $f$-electrons is a key question in certain actinide compounds.

URu$_2$Si$_2$ is such a system, known for a phase transition at 17.5 K ($T_{\rm{HO}}$) into the ``hidden order" (HO) state \cite{Maple_1986}. Despite decades of intensive research \cite{mydosh2011colloquium}, whether the HO is primarily associated with itinerant \cite{bareille2014momentum,hall2014optical,bachar2016detailed} or localized electrons \cite{park2012observation,park2014hidden,bourdarot2014neutron,kung2015chirality} remains to be unambiguously determined. According to previous studies by some of us using quasiparticle scattering spectroscopy (QPS) \cite{park2012observation,park2014hidden}, the hybridization gap opens well above $T_{\rm{HO}}$, questioning the hybridization process being directly responsible for the HO \cite{schmidt2010imaging,dubi2011hybridization,riseborough2012phase,chandra2013hastatic}. This result also poses a challenge to the Fermi surface (FS) gapping picture, widely adopted to explain gap-like behaviors \cite{oh2007interplay,bareille2014momentum,bachar2016detailed,ran2016phase}. This is because the corresponding drastic change in the spectral density must be detected by QPS \cite{nowack1992calculation,naidyuk2005point} as it exploits ballistic transport near the Fermi level \cite{naidyuk2005point,park2009andreev}, but no such signature was actually observed \cite{park2012observation,park2014hidden}. Another remaining issue is that the gap values extracted from different measurements are somewhat discrepant, e.g., in an analysis of electrical resistivity, heat capacity, and thermal expansion coefficient data \cite{ran2016phase}. The electrical resistivity has been frequently fit to expressions derived for the scattering off magnon-like excitations \cite{Palstra_1986,mentink1996gap,ran2016phase}. However, despite the likely existence of such collective modes, associating the extracted gap with the FS is questionable since it must be for the spin, rather than charge, sector. In addition, the resistivity jump at $T_{\rm{HO}}$, taken widely as a strong evidence for carrier depletion upon the FS gapping, needs to be explained quantitatively. After all, gap-like behaviors in {U}{R}u$_2${S}i$_2$ may reflect different aspects of the HO problem rather than sharing a single cause; thus, it is crucial to distinguish their origins.

\begin{figure}[tbp]
 \includegraphics[width=1\linewidth]{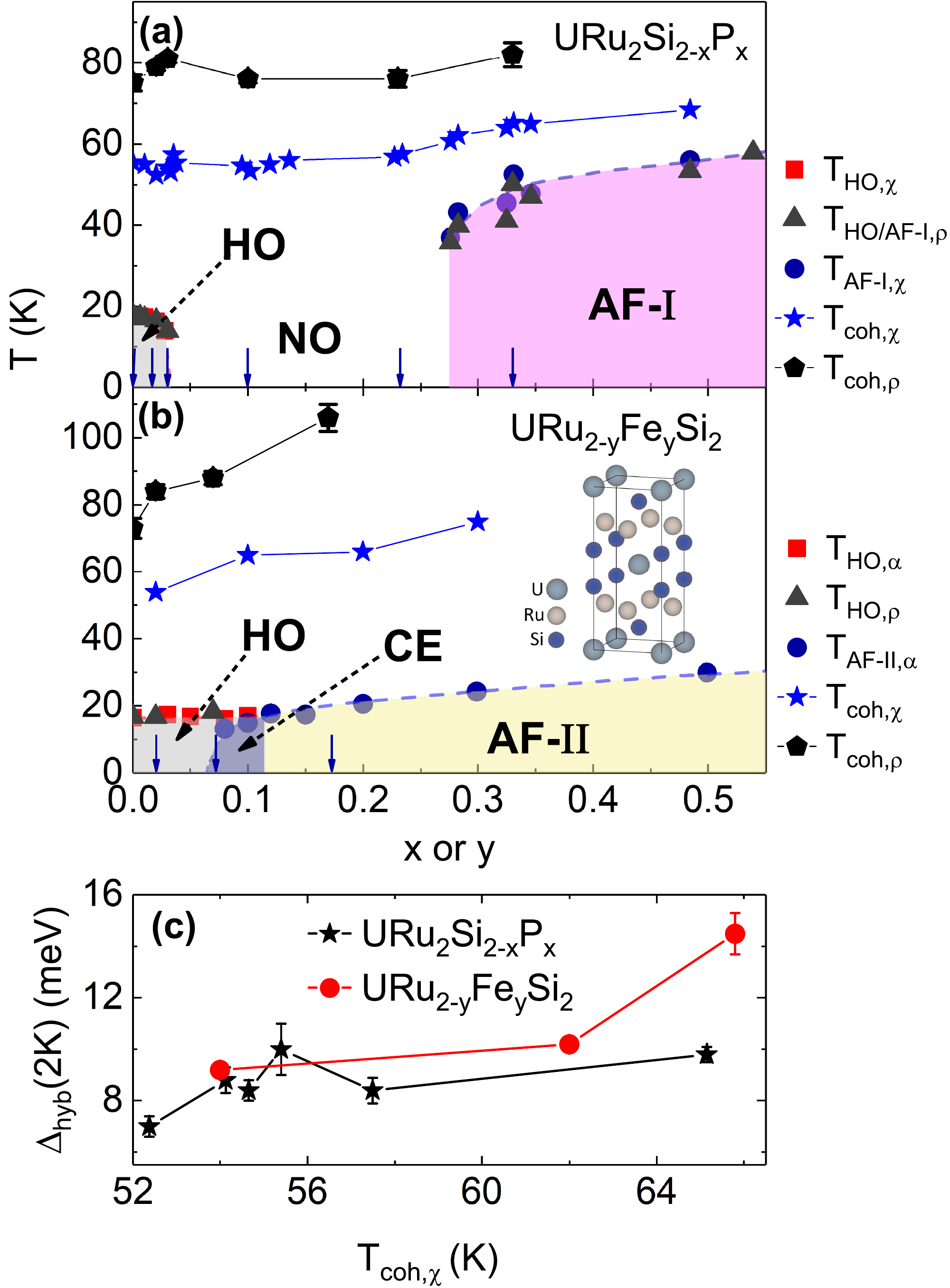}
 \caption{\small{(a) Temperature vs. P-content (T-x) phase diagram of URu$_2$Si$_{\rm{2-x}}$P$_{\rm{x}}$, adapted from Ref.~\citenum{Gallagher_2017} and based on the measurements of magnetic susceptibility ($\chi$) and electrical resistivity ($\rho$). AF-I stands for antiferromagnetic order and NO for no-order. $T_{\rm{coh}}$ denotes the coherence temperature. (b) T-$y$ phase diagram of URu$_{\rm{2-y}}$Fe$_{\rm{y}}$Si$_2$  constructed based on thermal expansion coefficient($\alpha$) and resistivity \cite{ran2016phase}, and magnetic susceptibility measurements \cite{wilson2016antiferromagnetism}. AF-II stands for antiferromagnetic order and CE for coexisting orders. The inset depicts a unit cell of URu$_2$Si$_2$. In both panels, vertical arrows along the horizontal axis indicate substituent concentrations studied in this work.} (c) Hybridization gap ($\Delta_{\rm{hyb}}$) at $T$ = 2 K (from Figs. 2 \& 3) vs. coherence temperature ($T_{\rm{coh,\chi}}$). The solid lines are guide to the eyes.}
 \label{fig:PD}
 \end{figure}

Another approach is to investigate how the HO is related to other phases induced by tuning quantum critical control parameters. The effect of chemical substitution has been extensively studied including Rh \cite{kawarazaki1994frozen,yokoyama2004neutron} and Os \cite{kanchanavatee2014enhancement}. In particular, P in URu$_2$Si$_{\rm{2-x}}$P$_{\rm{x}}$ nominally adds conduction electrons. However, unlike most other substituents, the phase diagram spans a no-order (NO, i.e., paramagnetic) region that separates the HO completely from an antiferromagnetic phase (AF-I) \cite{gallagher2016unfolding,Gallagher_2017}, as shown in Fig.~\ref{fig:PD} (a). On the other hand, the isoelectronic substitution of Fe in URu$_{\rm{2-y}}$Fe$_{\rm{y}}$Si$_2$ causes a continuous transformation of the HO into another antiferromagnetic phase (AF-II) with a co-existing (CE) region in between \cite{ran2016phase}, as shown in Fig.~\ref{fig:PD} (b). This phase diagram closely resembles that of the parent compound under pressure \cite{hassinger2008temperature},  suggesting the Fe substitution effectively acts like applying hydrostatic pressure \cite{wolowiec2016evolution}. The pressure-induced large moment antiferromagnet (LMAF) was found to return to the HO under a strong magnetic field \cite{aoki2009field}. The smooth evolution between the HO and AF-II or LMAF suggests their underlying interactions may be rooted on the same ingredients, unlike for AF-I. Thus, comparative studies of all these phases should bring novel insights into the HO problem.

In this Rapid Communication, we report a combined study via QPS and resistivity measurements over the entire phase space for URu$_2$Si$_{\rm{2-x}}$P$_{\rm{x}}$ and URu$_{\rm{2-y}}$Fe$_{\rm{y}}$Si$_2$. The hybridization gap is observed in all phases including the NO region and evolves smoothly across phase boundaries, indicating the hybridization is a general process for heavy fermions rather than a driving force for phase transitions. For a consistent understanding of all gap-like behaviors, we advance a novel interpretation of the electrical resistivity by considering the scattering off gapped bosonic excitations in the ordered state \cite{jobiliong2005magnetization}. By allowing an unusual temperature dependence of their \textit{stiffness}, the entire characteristics including the jump can be nicely replicated. Our analysis also reveals the different nature of the AF-I from AF-II phases, for whose microscopic understanding we provide speculations on how differently P and Fe substitutions affect the underlying interactions.

URu$_2$Si$_{\rm{2-x}}$P$_{\rm{x}}$ and URu$_{\rm{2-y}}$Fe$_{\rm{y}}$Si$_2$ single crystals were grown by molten metal flux \cite{gallagher2016unfolding} and Czochralski methods \cite{ran2016phase}, respectively. The (0 0 1) surface of the crystals was then manually polished to a few nm peak-to-dip roughness. Such a smooth surface is essential in making a spectroscopic junction (Sect. I in Ref.~\citenum{supplementary}) free from local heating effect that obscures intrinsic information (Sect. II in Ref.~\citenum{supplementary}). QPS junctions were formed using Au tips \cite{narasiwodeyar2015two} in a custom-built rig \cite{tortello2016design} and differential conductance across the junction was measured using a standard four-probe lock-in technique. The conductance data were analyzed using the Maltseva-Dzero-Coleman (MDC) model (\cite{maltseva2009electron}, also see Sect. III in Ref.~\citenum{supplementary}), according to which the conductance curve can be asymmetric due to a Fano resonance \cite{fano1961u} between the two \textit{co-tunneling} channels into a Kondo lattice. DC electrical resistance was measured with the four-probe method and analyzed using a model proposed by Jobiliong \etal \cite{jobiliong2005magnetization}.

\begin{figure}[bpt]
 \includegraphics[width=1\linewidth]{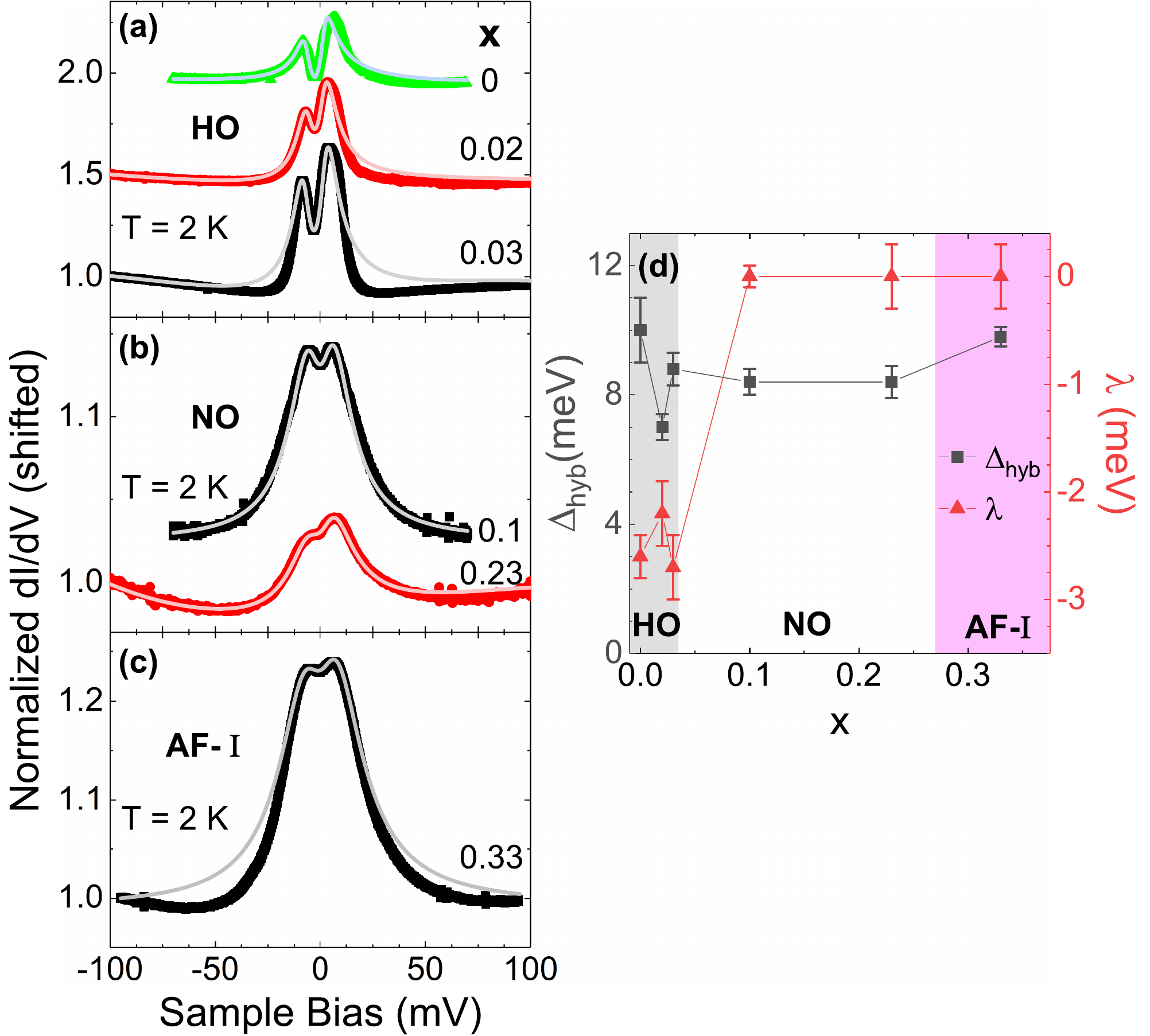}%
 \caption{\small{(a)-(c) Normalized differential conductance taken from junctions on URu$_2$Si$_{\rm{2-x}}$P$_{\rm{x}}$ at T = 2 K (colored symbols) and best fit curves (solid gray lines). In (a) \& (b), data and fit curves are shifted vertically for clarity. (d) Hybridization gap ($\Delta_{\rm{hyb}}$) and renormalized $f$-level ($\lambda$) extracted from an analysis using the MDC model. Labels denote different regions in the phase diagram.}}
 \label{fig:URSP_QPS}
 \end{figure}

The conductance spectra for URu$_2$Si$_{\rm{2-x}}$P$_{\rm{x}}$ with different P content (x) are displayed in Figs.~\ref{fig:URSP_QPS} (a)-(c). They all exhibit an asymmetric double-peak structure resulting from the above-mentioned Fano resonance in a Kondo lattice \cite{maltseva2009electron}. It becomes smeared at large x as the electronic mean free path gets shortened due to increasing disorder, also reflected in the corresponding decrease of the residual resistivity ratio (RRR) (Sect. II in Ref.~\citenum{supplementary}). This structure has been established as signifying an indirect gap in the hybridized bands through recent theoretical \cite{maltseva2009electron,fogelstrom2010point,figgins2010differential,wolfle2010tunneling} and experimental \cite{park2008andreev,park2009andreev,park2012observation, park2014hidden,park2014hybridization,jaggi2017hybridization} studies. The solid lines are best fit curves using the modified MDC model (\cite{maltseva2009electron,wolfle2010tunneling}, also Sect. III in Ref.~\citenum{supplementary}). The extracted hybridization gap ($\Delta_{\rm{hyb}}$) and renormalized $f$-level ($\lambda$) are shown as a function of x in Fig.~\ref{fig:URSP_QPS} (d). The gap size for the parent compound is about 10 meV, similar to the values obtained from the previous QPS \cite{park2012observation,park2014hidden} and recent optical conductivity measurements \cite{hall2014optical,bachar2016detailed}. Note $\Delta_{\rm{hyb}}$ changes very little as a function of x, similarly to $T_{\rm{coh}}$ vs. x shown in Fig. 1(a) and, thus, roughly conforming to the known correlation, $\Delta_{\rm{hyb}} \propto T_{\rm{coh,\chi}}$ \cite{dordevic2001hybridization}, as plotted in Fig.~\ref{fig:PD} (c). In particular, a hybridization gap is still observed in the NO region, similar in magnitude to those in the HO and AF-I regions, clearly indicating it is not associated with emergent ordering.

\begin{figure}[t]
 \includegraphics[width=1\linewidth]{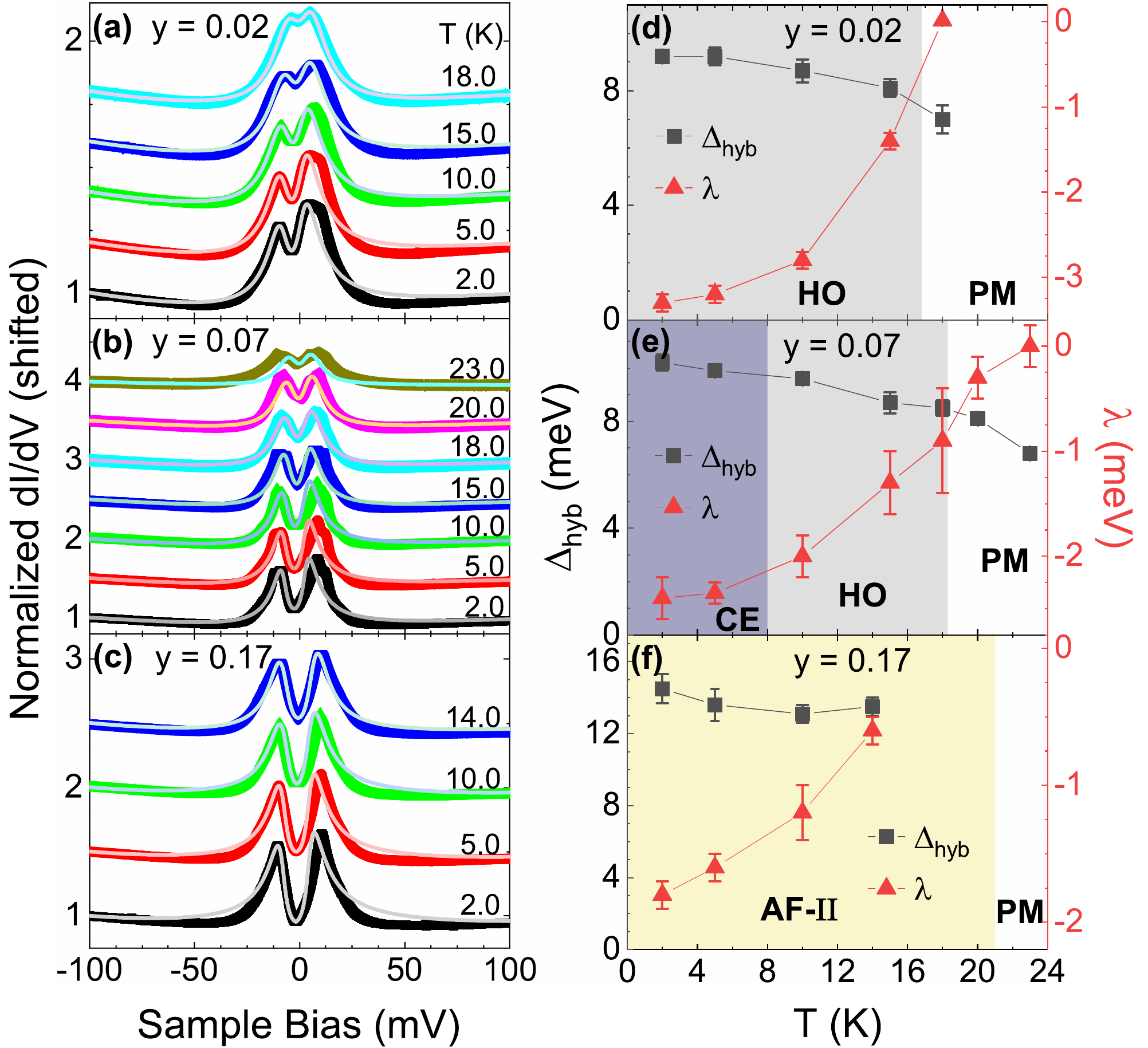}%
 \caption{\small{(a)-(c) Temperature-dependent conductance spectra taken from junctions on URu$_{\rm{2-y}}$Fe$_{\rm{y}}$Si$_2$ (colored symbols) and best fit curves (solid gray lines). Data and fit curves are shifted vertically for clarity. (d)-(f) Extracted hybridization gap ($\Delta_{\rm{hyb}}$) and renormalized $f$-level ($\lambda$).}}
 \label{fig:URFS_QPS}
 \end{figure}

Temperature-dependent conductance spectra for URu$_{\rm{2-y}}$Fe$_{\rm{y}}$Si$_2$ are shown in Figs.~\ref{fig:URFS_QPS} (a)-(c). Again, a hybridization gap is observed in all phases. In contrast to URu$_2$Si$_{\rm{2-x}}$P$_{\rm{x}}$, the conductance curve doesn't exhibit noticeable change in the sharpness with increasing Fe content (y), in agreement with the RRR not changing much. In the HO and CE regions, the hybridization gap opens well above $T_{\rm{HO}}$, in agreement with recent optical conductivity measurements of the parent compound \cite{hall2014optical,bachar2016detailed}. In the AF-II region, the junction became unstable above 14 K but the sharpness of the double-peak structure implies that the hybridization gap may remain open well above $T_{\rm{AF-II}}$. The temperature dependence of the hybridization gap and the renormalized $f$-level are plotted in Figs.~\ref{fig:URFS_QPS} (d)-(f). Like in URu$_2$Si$_{\rm{2-x}}$P$_{\rm{x}}$, the gap at 2 K roughly exhibits the correlation, $\Delta_{\rm{hyb}} \propto T_{\rm{coh,\chi}}$ \cite{dordevic2001hybridization}, as shown in Fig.~\ref{fig:PD} (c). With increasing temperature, in all three regions, $\Delta_{\rm{hyb}}$ decreases and $\lambda$ approaches zero (the Fermi level), similar to what occurs in the parent compound \cite{park2012observation}. Furthermore, all properties including the conductance shape, $\lambda$, and $\Delta_{\rm{hyb}}$ exhibit a smooth evolution without any anomaly upon crossing the phase transition temperature. While our result doesn't rule out a reconstruction of the FS at $T_{\rm{HO}}$, it can't be understood within the FS gapping picture \cite{bareille2014momentum,bachar2016detailed}. Meanwhile, upon suppressing the HO in both URu$_2$Si$_{\rm{2-x}}$P$_{\rm{x}}$ and URu$_{\rm{2-y}}$Fe$_{\rm{y}}$Si$_2$ (by chemical substitution or temperature), $\lambda$ goes from negative to zero, as shown in Fig.~\ref{fig:URSP_QPS} (d) and Figs.~\ref{fig:URFS_QPS} (d)-(f). Inferring from the well-known single impurity Kondo resonance of width \rm{W}, for which the resonance energy is expressed as  $\varepsilon_0 = \frac{\rm{W}}{2} \tan[(1 - n_f)\frac{\pi}{2}]$ \cite{hewson1997kondo}, the above-described behavior of $\lambda$  may indicate an accompanying change in the $f$-level occupancy ($n_f$). This speculation is in line with a recent proposal invoking a possible valence change associated with the HO transition \cite{harrison2019hidden}. In addition, according to resonant X-ray emission and electron-energy loss spectroscopy measurements on {U}{R}u$_2${S}i$_2$ \cite{jeffries2010degree,booth2016probing}, the 5$f$ electron count in the HO is quite far away from an integer, in agreement with our $\lambda$ being finite in the HO region. It is also notable that observations similar to ours for $\lambda$ and $\Delta_{\rm{hyb}}$ in URu$_{\rm{2-y}}$Fe$_{\rm{y}}$Si$_2$ have been reported in a recent photoemission study on {U}{R}u$_2${S}i$_2$ \cite{boariu2013momentum}: a $\Pi$-shaped quasiparticle band at the $\Gamma$-point shifts to the Fermi level from below with increasing temperature, whereas the hybridization gap at the $X$-point doesn't change.


Previous QPS studies on {U}{R}u$_2${S}i$_2$ concluded that the hybridization gap is not the HO order parameter \cite{park2012observation,park2014hidden}. Furthermore, a hybridization gap is observed to open well above the N{\'e}el temperature in another related U-based compound, UPd$_2$Al$_3$, known to be a local-moment antiferromagnet \cite{caspary1993unusual}. In the present study, the opening of a hybridization gap even in the NO region corroborates that hybridization is a generic process underlying the formation of heavy fermion bands and thus, is not correlated with the type of an emergent order. As speculated previously \cite{park2012observation,park2014hidden}, our observation questions the plausibility of the FS gapping scenario. Related to this, it is important to note that while the FS topology undergoes no significant change in the HO-to-LMAF transition \cite{ohkuni1999fermi,hassinger2010similarity}, the magnetic moment becomes finite abruptly upon the transition \cite{niklowitz2010parasitic}. Within the itinerant picture, the LMAF must arise from FS nesting. In turn, the abrupt increase in magnetic moment should reflect a large change in the FS topology, in apparent contradiction with quantum oscillation results \cite{ohkuni1999fermi,hassinger2010similarity}. This suggests that both the HO and LMAF are more likely associated with localized electrons rather than itinerant electrons.

\begin{figure*}[btp]
 \includegraphics[width=1\textwidth]{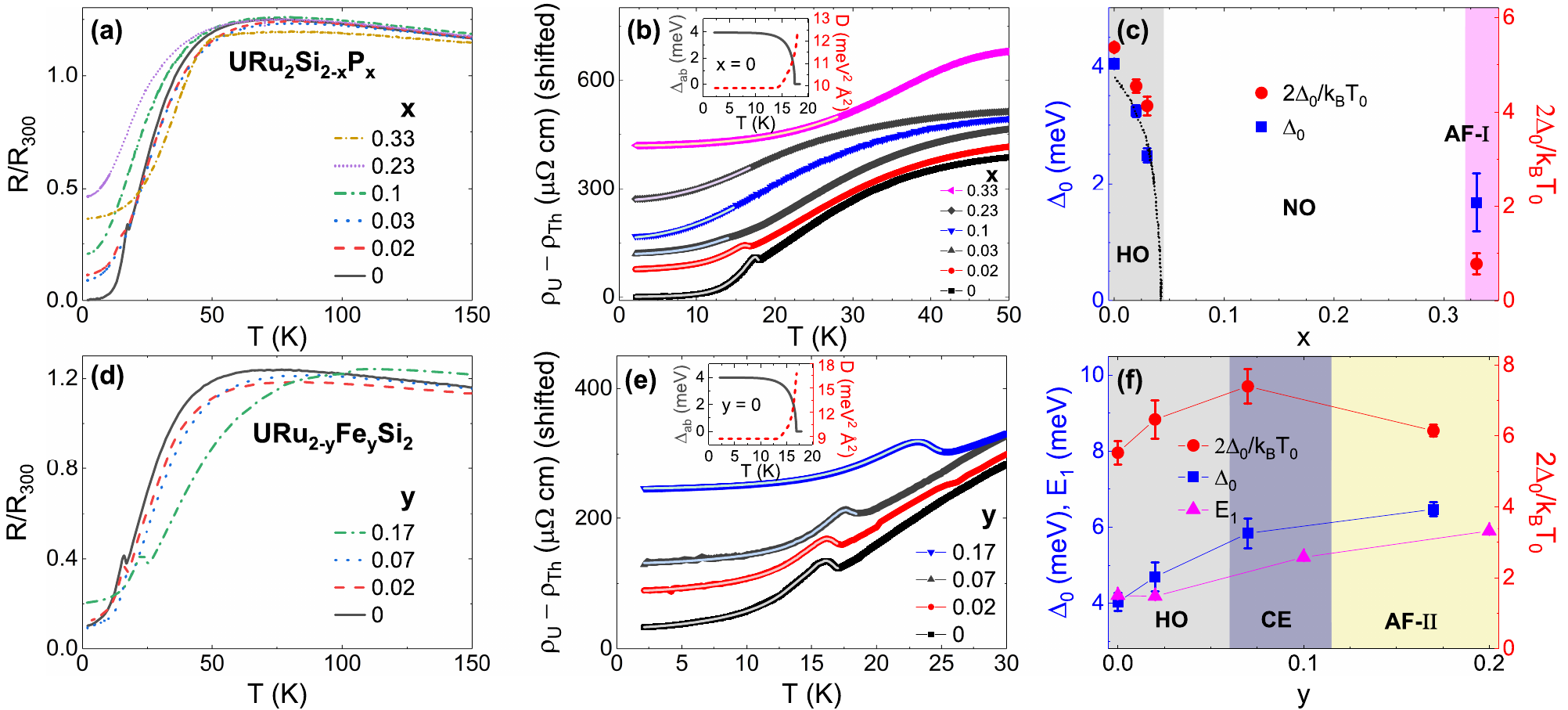}%
 \caption{\small{(a) Normalized resistance vs. temperature for URu$_2$Si$_{\rm{2-x}}$P$_{\rm{x}}$. (b) Resistivity in the low-temperature region with the phonon contribution subtracted out (colored symbols). $\rho_{\rm{U}}$ is the resistivity of URu$_2$Si$_{\rm{2-x}}$P$_{\rm{x}}$ and $\rho_{\rm{Th}}$ represents the resistivity of ThRu$_2$Si$_2$. Data and fit curves are shifted vertically for clarity. The inset shows temperature-dependent bosonic excitation gap (left-axis, solid line) and stiffness (right-axis, dashed line) that are extracted for the parent compound. (c) Blue squares indicate the extracted gap value at zero temperature. The dashed line is a guide to the eye. Red circles represent the gap ratio, 2$\Delta_0/k_BT_0$, where $T_0$ is the ordering temperature. (d)-(f) Same as (a)-(c) but for URu$_{\rm{2-y}}$Fe$_{\rm{y}}$Si$_2$. The E$_1$ gap for URu$_{\rm{2-y}}$Fe$_{\rm{y}}$Si$_2$ in (f) is from Ref.~\citenum{williams2017hidden}. Solid lines are a guide to the eye.}}
 \label{fig:resis}
 \end{figure*}

Within such a localized picture, some of us previously showed \cite{park2012observation} that the gap-like behavior of the in-plane electrical resistivity in {U}{R}u$_2${S}i$_2$ is associated with the E$_1$ gap detected in inelastic neutron scattering (INS) by analyzing the resistivity with a model proposed by Jobiliong \textit{et al.} \cite{jobiliong2005magnetization}. This model explains the temperature dependence of resistivity in antiferromagnets in terms of scattering off gapped magnon excitations. In the previous analysis, similar gapped bosonic excitations were assumed to exist in the HO, and it was shown the entire resistivity curve, including the jump at $T_{\rm{HO}}$, could be fit by this model. But the expression used in this fit is valid only in the low-temperature limit (T $\ll \Delta$). Thus, here (Sect. IV in Ref.~\citenum{supplementary}) we use a more general expression that is not subject to this constraint \cite{jobiliong2005magnetization}:
\begin{gather}
	\rho(T) = \rho_0 + AT^2+ \nonumber \\
    \frac{B}{T} \int_{0}^{\infty} \frac{k^4\sqrt{\Delta_{\rm{ab}}(T)^2 + D(T)k^2}}{\sinh^2(\sqrt{\Delta_{\rm{ab}}(T)^2+D(T) k^2}/2T)} dk,
    \label{eq:Jobilliong}
\end{gather}
where $\rho_0$ is the residual resistivity and the second term describes the Fermi liquid behavior. Scattering off the bosonic excitations is accounted for by the third term, where $k$ is the wave number, $D(T)$ is the stiffness, and the gap $\Delta_{\rm{ab}}(T) = \Delta_0 \tanh(3.2 \sqrt{T_0/T - 1})$. $\Delta_0$ is the zero-temperature gap and $T_0$ is the ordering temperature. For more accurate estimation of the gap, the phonon contribution is eliminated by subtracting out the resistivity of ThRu$_2$Si$_2$, a compound iso-structural to {U}{R}u$_2${S}i$_2$. By setting the stiffness as a free parameter, the entire resistivity curve including the jump at $T_0$ can be reproduced. Best fit curves are shown in Figs.~\ref{fig:resis} (b) and (e) with $\Delta_{\rm{ab}}$ and $D$ plotted in the insets for the parent compound for illustration (Sect. IV in Ref.~\citenum{supplementary}). While the data for AF-I with no jump at $T_{\rm{AF-I}}$ were analyzed using the same approximate expression as in Ref.~\citenum{park2012observation}, the data for the HO, CE, and AF-II regions are all nicely fit by this general expression. Notably, the data for the NO region is fit well by an expression containing only the first two terms in Eq.~\ref{eq:Jobilliong} (Sect. IV in Ref.~\citenum{supplementary}), in agreement with the disappearance of bosonic excitations in this region.

The extracted zero-temperature gaps are plotted as a function of substituent concentration in Figs.~\ref{fig:resis} (c) and (f). The gap ratio, 2$\Delta_0/k_{\rm{B}}T_0$, ranges from 4 - 7.5 except for AF-I, for which it is only $\sim$ 0.75, much smaller than 3.53 from the weak-coupling mean field theory. This indicates that AF-I is of different nature from AF-II, as speculated earlier. A similarly small gap-ratio ($\sim$ 2) and the same kink-decay (instead of jump-decay) behavior are also observed in UPd$_2$Al$_3$ \cite{jaggi2017hybridization}, suggesting that AF-I is more likely due to local moments, consistent with a recent nuclear magnetic resonance study \cite{shirer2017nmr}. The strongly contrasting properties between AF-I and AF-II might be due to the different roles played by different chemical substitutions, as mentioned earlier. As shown in the inset of Fig.~\ref{fig:PD} (b), the Si sites are closer to the U sites than the Ru sites. Therefore, P substitution may affect the interaction that is responsible for the HO more drastically, transforming it into a rather conventional antiferromagnetic interaction. It is also notable that with increasing Fe-content, the extracted gap closely follows the INS E$_1$ gap at \textbf{Q}$_1$ = (1.4, 0, 0) \cite{williams2017hidden}, as shown in Fig.~\ref{fig:resis} (f), indicating that the INS E$_1$ gap may originate from the same gapped bosonic excitations as for the resistivity. Such association is also supported by the similar temperature dependence of the two gaps \cite{bourdarot2014neutron}. There is another gap detected by INS at \textbf{Q}$_0$ = (1, 0, 0). However, this gap is not only much smaller (E$_0$ = 1.7 meV for y = 0) but also detected only in the HO region \cite{bourdarot2014neutron}, so it is unlikely to play a significant role in the resistivity jump-decay behavior. In contrast to the kink-decay behavior in the AF-I state, both HO and AF-II have the same jump-decay behavior and the bosonic excitation gap increases continuously when going from HO to AF-II, closely following $T_0$. These observations suggest that HO and AF-II may share a common order parameter. Such a model has been put forward recently to explain a resonance mode in Raman scattering observed in both the HO and AF-II phases \cite{kung2016analogy}. Accordingly, one can imagine that similar bosonic excitations may exist in both phases, in line with our findings. Our analysis should also be applicable to gap-like behaviors in other experiments. For example, with decreasing temperature, the Hall coefficient in {U}{R}u$_2${S}i$_2$ abruptly increases at $T_{\rm{HO}}$ then decays slowly at lower temperature, and this behavior was attributed to the depletion of charge carriers due to FS gapping \cite{kasahara2007exotic}. Instead, the dominant scattering off gapped bosonic excitations in the HO or AF-II, in combination with a similarly anomalous temperature dependence of their stiffness, can qualitatively explain this behavior, similar to the case of SrRuO$_3$ where magnons are known to play a key role \cite{jenni2019interplay}.

According to the working principle for QPS, such scattering off gapped bosonic excitations would show up as weak non-linearity in the current-voltage characteristics at a bias voltage corresponding to the gap ($\sim$ 4 meV) \cite{naidyuk2005point}. Such a signature is not detected in our measurements presumably because it is buried in the conductance that varies rapidly due to the hybridization gap. It could be revealed in a second harmonic measurement, analogously to phonons in simple metals \cite{naidyuk2005point}.

In conclusion, our QPS study on URu$_2$Si$_2$ containing P and Fe substituents reveals that a hybridization gap opens regardless of the emergent ordering including the NO without any anomaly upon crossing the phase boundary, indicating the hybridization is a general process instead of driving the phase transition. Because QPS detects quasiparticle scattering near the Fermi level, this result suggests the HO originates from localized (rather than itinerant) electrons. For a comprehensive understanding of all gap-like behaviors, we advance a new analysis of the electrical resistivity based on the scattering off gapped bosonic excitations, accounting for all of the characteristics, including the jump at the transition. The extracted gap is in agreement with the E$_1$ gap in INS. A similar approach can also provide a natural explanation for the Hall effect. Our results suggest the multitude of $f$-electrons in {U}{R}u$_2${S}i$_2$ may play intriguing roles leading to intertwined orders (HO and AF), whose analogs can be found in other correlated systems \cite{stewart2011superconductivity,fradkin2015colloquium,kim2016intertwined}.

This work was supported by the US National Science Foundation (NSF), Division of Materials Research (DMR) under Grant No. NSF/DMR-1704712 (S.Z., L.H.G., and W.K.P.). The work at the National High Magnetic Field Laboratory is partly supported by the NSF/DMR-1644779 and the state of Florida. The research at UC San Diego was supported by the US Department of Energy, Office of Basic Energy Sciences, Division of Materials Sciences and Engineering, under Grant No. DEFG02-04-ER46105 (single crystal growth) and US NSF/DMR-1206553 (physical properties measurements).

\bibliography{Origin_of_the_gaplike_behaviors_in_URS}
\end{document}



\title{Supplemental Material: Origins for gap-like behaviors in {U}{R}u$_2${S}i$_2$ -- a combined study via quasiparticle scattering spectroscopy and resistivity measurements}


\author{S. Zhang$^{1,2}$, G. Chappell$^{1,2}$, N. Pouse$^3$, R. E. Baumbach$^{1,2}$, M. B. Maple$^3$, L. H. Greene$^{1,2}$, W. K. Park$^{1,}$}
\email{wkpark@magnet.fsu.edu}
\affiliation{$^1$National High Magnetic Field Laboratory, Florida State University, Tallahassee, Florida 32310\\
$^2$Department of Physics, Florida State University, Tallahassee, Florida 32306\\
$^3$Department of Physics, University of California, San Diego, California 92093}

\date{\today}



\pacs{}

\maketitle

\setcounter{equation}{0}
\setcounter{figure}{0}
\setcounter{table}{0}
\setcounter{page}{1}
\makeatletter
\renewcommand{\theequation}{S\arabic{equation}}
\renewcommand{\thefigure}{S\arabic{figure}}
\renewcommand{\thetable}{S\arabic{table}}
\renewcommand{\etal}{\textit{et al. }}
\renewcommand{\insitu}{\textit{in situ}}

\section{I. Experiments}
As mentioned in the main text, URu$_2$Si$_2$ single crystals containing P and Fe substituents were grown using molten metal flux \cite{gallagher2016unfolding} and Czochralski techniques \cite{ran2016phase}, respectively. During a flux growth, crystals in the URu$_2$Si$_2$ family usually grow along the ab-plane, forming plate-like shape. In contrast, the major axes of a crystal grown by the Czochralski method cannot be identified easily from the shape. Therefore, single crystal X-ray diffractometry was used to find the (0 0 1) surface of the Fe-substituted URu$_2$Si$_2$ crystals. For both substitution series, the chemical composition was determined by energy-dispersive X-ray spectroscopy. Crystals were then mounted onto epoxy molds with the (0 0 1) surface exposed to the air. This surface often has step-like structures. Although the areas within each step can be flat, polishing was still desirable to produce a cleaner and smoother surface over a larger area, where a point-contact junction was formed using an Au tip. Finally, differential conductance was measured across the junction and the conductance spectra were analyzed as described in the main text and in Sect. III. To check the reproducibility, we made multiple junctions \insitu~by using a stack of two piezo nano-positioners \cite{tortello2016design}.

\begin{figure}[b]
 \includegraphics[width=\linewidth]{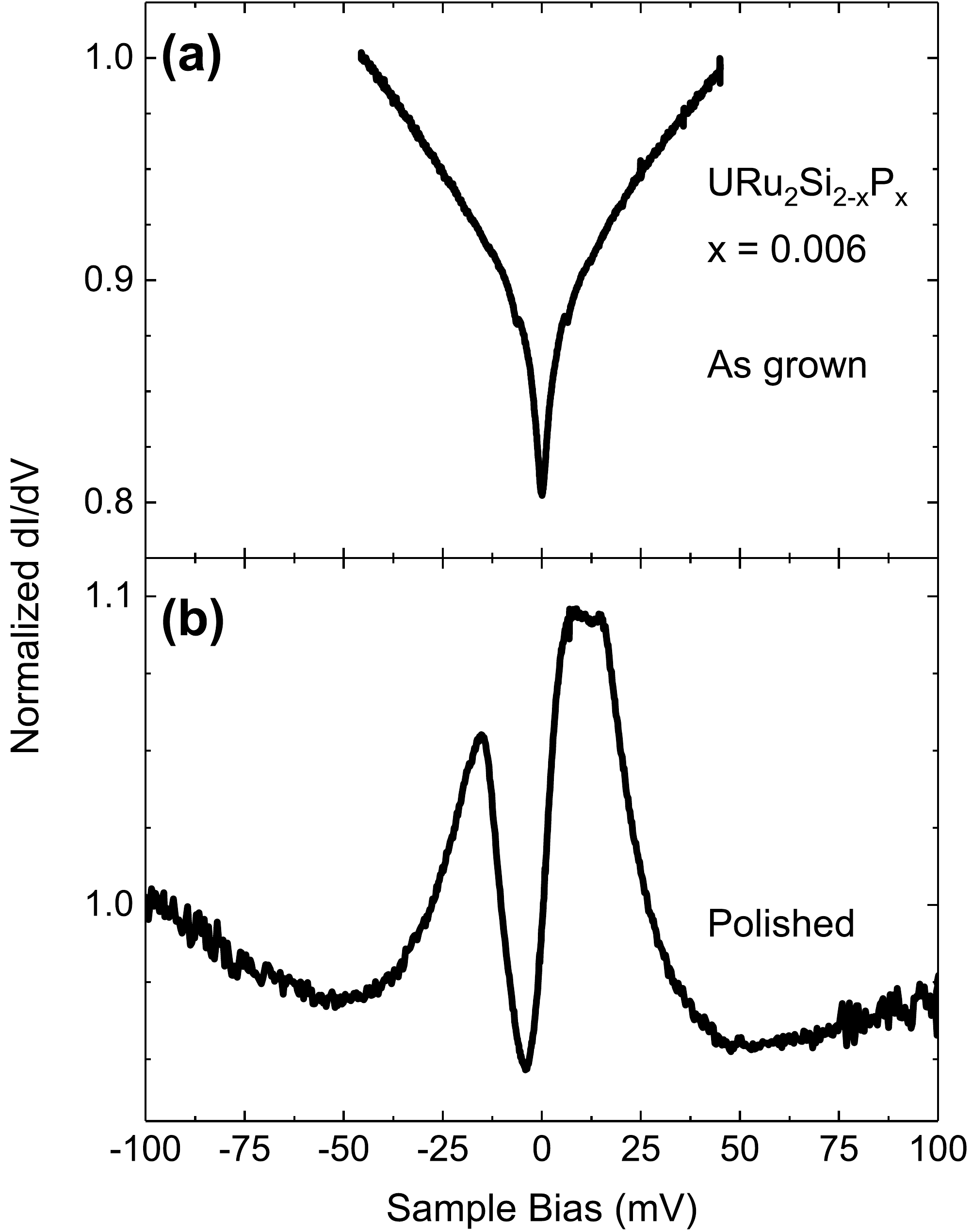}%
 \caption{(a) Differential conductance for an as-grown URu$_2$Si$_{2-x}$P$_x$ crystal with x = 0.006. (b) Conductance data for the same crystal taken after polishing. Data were taken at 2.0 K.}
 \label{fig:QPSpolish}
\end{figure}

\begin{figure}[tbp]
 \includegraphics[width=0.9\linewidth]{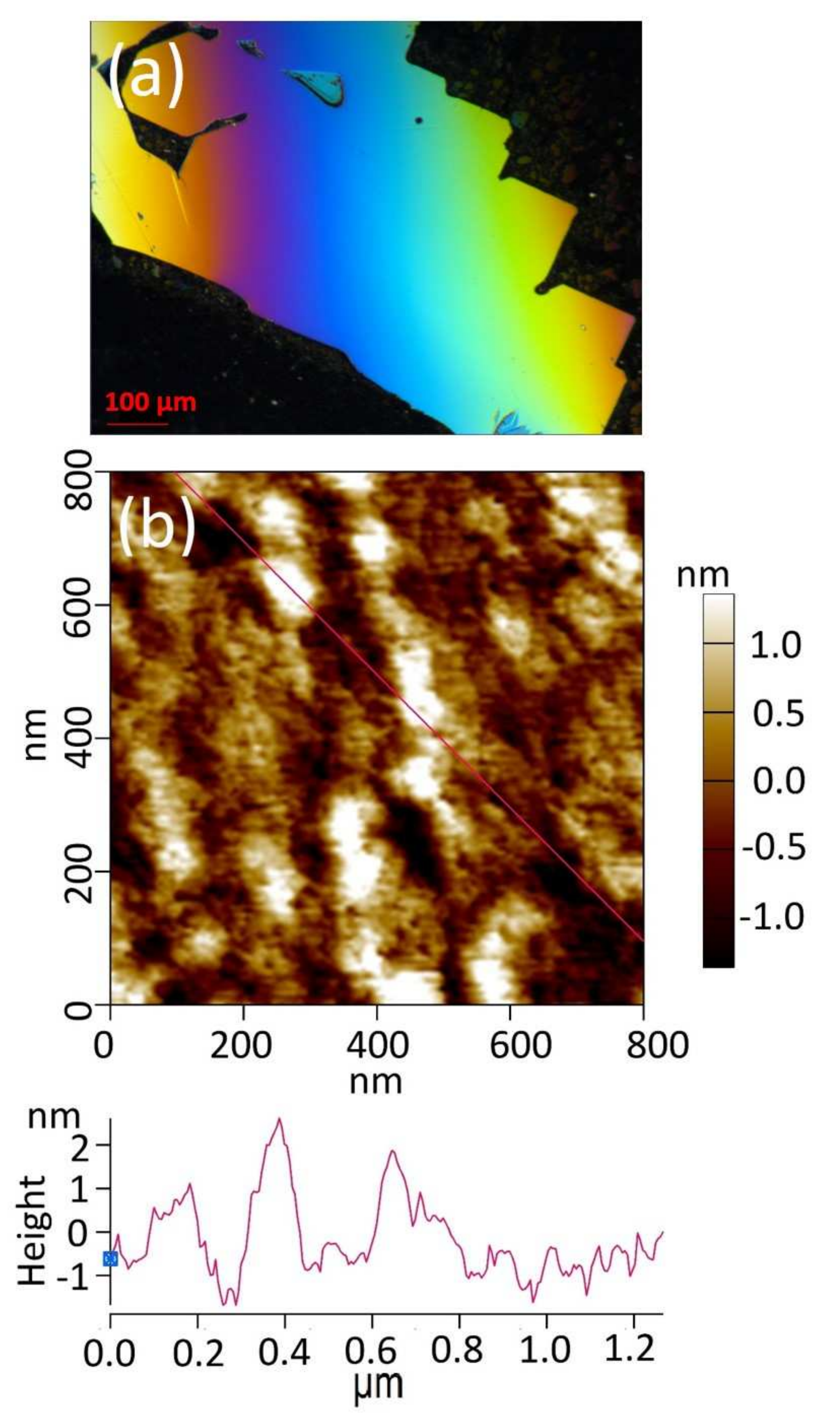}%
 \caption{(a) Typical optical microscope image of a polished URu$_2$Si$_2$ crystal. The color gradient is due to tilting of the sample surface. (b) Top panel shows an AFM height-retrace image. Bottom panel shows quantitative roughness of the selected region indicated by the red line in the top panel. Roughness of 2 nm is achieved after polishing.}
 \label{fig:smoothness}
\end{figure}

In the above-mentioned procedure, polishing is a very important step. Its effect is shown by performing quasiparticle scattering spectroscopy (QPS) on the same crystal before and after polishing. As plotted in Fig.~\ref{fig:QPSpolish}, before polishing, instead of the asymmetric double-peak structure, as seen reproducibly in polished crystals, only a sharp dip at zero bias was observed. Such conductance shape is typically observed when the surface is dirty. In addition, the junction was not stable over thermal cycling. After polishing, the asymmetric double-peak structure was observed and the junction became more stable. Images of a typical polished surface taken with an optical microscope and an atomic force microscope (AFM) are shown in Figs.~\ref{fig:smoothness} (a) and (b), respectively. No apparent scratch was detected and the surface was quite smooth with the maximum peak-to-dip distance of about 2-3 nm.

\section{II. QPS Diagnostics}
Since spectroscopic information can be readily obscured in QPS despite the technical simplicity of forming a junction, as illustrated above, interpretation of the conductance spectra has been heavily dependent on the junction quality. Therefore, appropriate diagnostics should be executed. The asymmetric double-peak structure frequently observed in URu$_2$Si$_2$ was originally interpreted to arise from a gap opened on the Fermi surface accompanying the hidden order (HO) transition \cite{hasselbach1992point,nowack1992calculation,escudero1994temperature, naidyuk1995anisotropy,samuely1995superconducting,thieme1995itinerant,steiner1996magnetic,rodrigo1997point,lu2012pressure}. However, through recent QPS studies by some of us \cite{park2012observation,park2014hidden}, the gaplike structure has been observed to occur well above the transition temperature, so a new interpretation was required. Its underlying process is now well established to be a lattice version of Fano resonance and its theoretical analysis using a model by Maltseva-Dzero-Coleman (MDC) \cite{maltseva2009electron} reveals spectroscopic information on the hybridization including the hybridization gap and renormalized $f$-level, as presented in the main text. The discrepancy in the gap opening temperature reported in different QPS studies \cite{hasselbach1992point,nowack1992calculation,escudero1994temperature, naidyuk1995anisotropy,samuely1995superconducting,thieme1995itinerant,steiner1996magnetic,rodrigo1997point,lu2012pressure} is also well understood as due to the different extent of junction's ballisticity (\cite{park2014hidden}, also see Supplemental Material in Ref.~\citenum{park2012observation}).  These findings should serve as a basis for diagnostics in a QPS study on URu$_2$Si$_2$, as detailed below.

\begin{figure}[tbp]
 \includegraphics[width=0.6\linewidth]{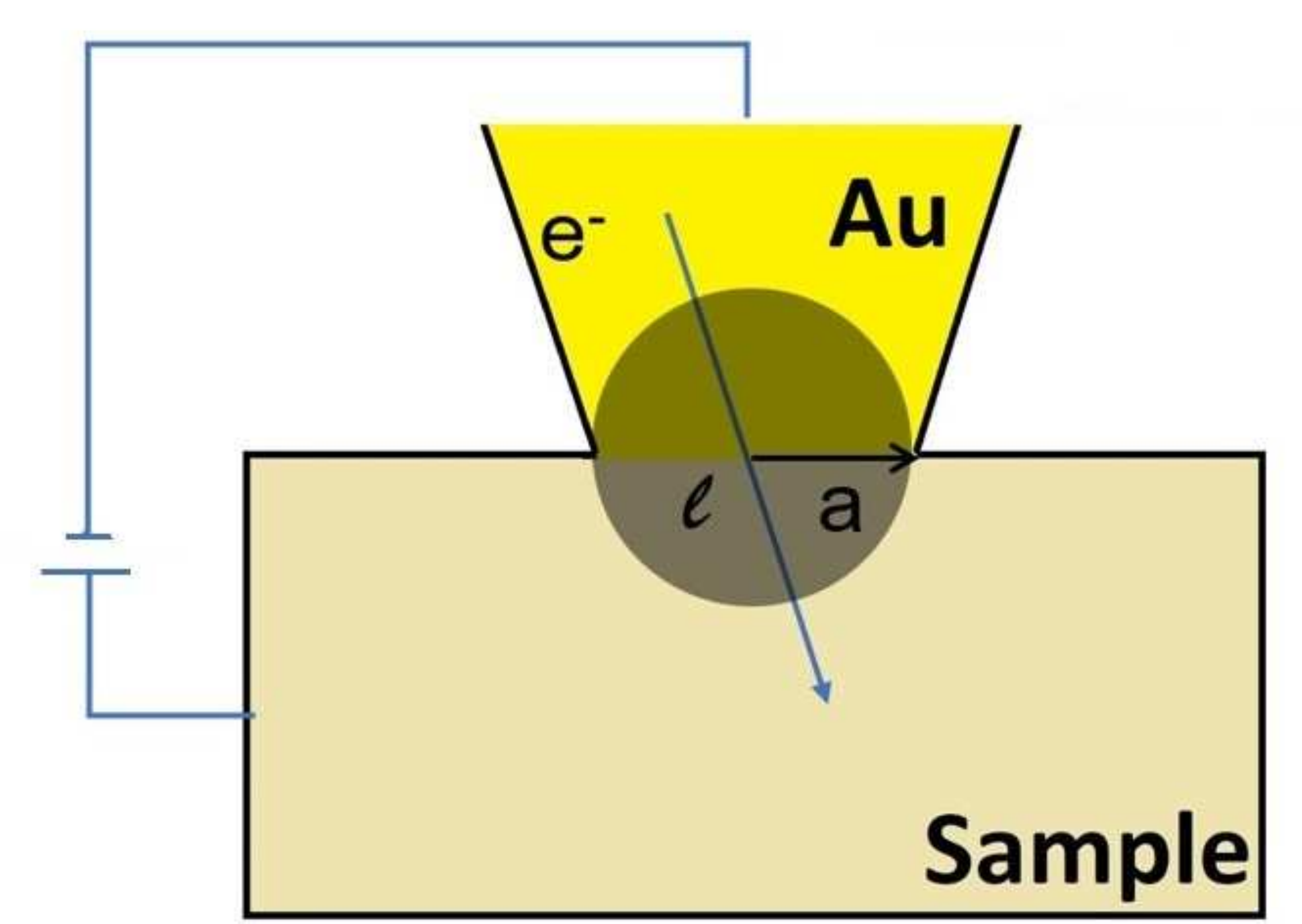}%
 \caption{Schematic drawing of a ballistic QPS junction where the electronic mean free path ($l$) is much longer than the junction size (2$a$), allowing quasiparticles to be injected into the sample without undergoing scattering within the junction area.}
 \label{fig:QPScartoon}
\end{figure}

\begin{figure}[tbp]
 \includegraphics[width=\linewidth]{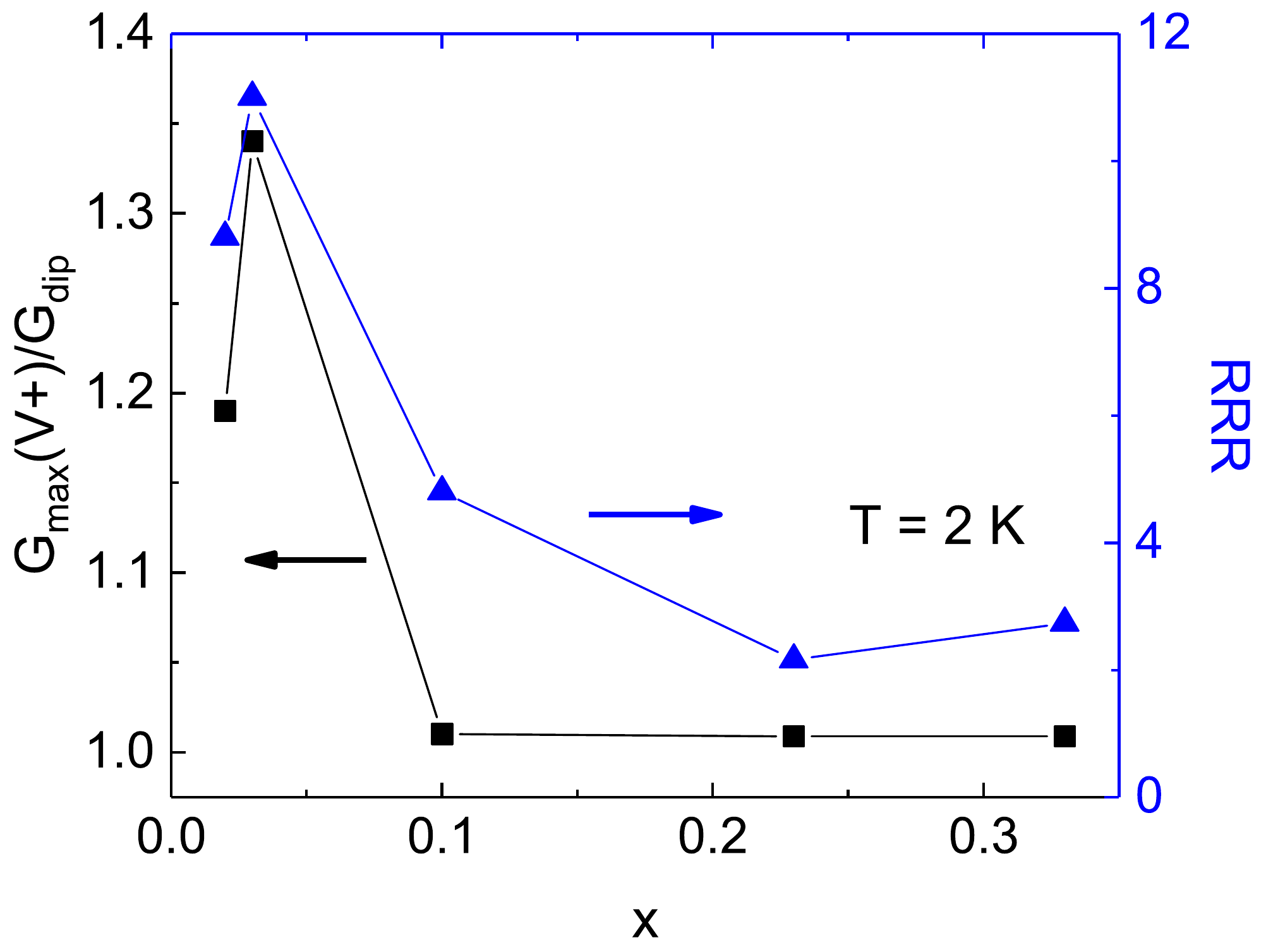}%
 \caption{Peak-to-dip ratio from QPS conductance spectra and RRR as a function of P-content for URu$_2$Si$_{2-x}$P$_x$. The conductance peak used for the ratio is the peak in the positive bias branch.}
 \label{fig:smearing}
\end{figure}

In a simplified view, a point-contact junction may fall into two regimes, spectroscopic or non-spectroscopic, based on its relative size with respect to electronic mean free paths (elastic and inelastic) \cite{park2009andreev,naidyuk2005point}. The spectroscopic regime consists of a ballistic regime and a diffusive regime. In the ballistic regime, quasiparticles are injected ballistically without undergoing scattering within the junction area since the junction is smaller than the $elastic$ mean free path, as illustrated in Fig.~\ref{fig:QPScartoon}, thus, avoiding the Joule heating. Their higher order scattering processes in the bulk results in non-linear current-voltage characteristics, which contains spectroscopic information on the scattering source. In the diffusive regime, before arriving in the bulk, quasiparticles scatter $elastically$ within the junction area, e.g., due to impurities, causing characteristic conductance features to be smeared but without the spectroscopic information lost completely. This smearing effect is seen in our conductance spectra obtained from URu$_2$Si$_{\rm{2-x}}$P$_{\rm{x}}$ crystals (see Fig. 2). Here, the double-peak structure is obviously less sharp in the NO region than in the HO region. More quantitatively, the peak-to-dip ratio in the conductance curve (G$_{max}$/G$_{dip}$) is found to follow the residual resistance ratio (RRR) as a function of the P content, as plotted in Fig.~\ref{fig:smearing}.  This indicates that the conductance smearing in URu$_2$Si$_{\rm{2-x}}$P$_{\rm{x}}$ is very likely due to increased scattering off the disorder that is introduced by P substitution.

Therefore, to make a stable junction in the spectroscopic regime, it is important to use an extremely sharp tip to minimize the junction size. Moreover, the sample surface should be very clean and smooth so the junction can be stable and the obscurities rooted on junction geometry can be avoided. In the current work, all Au tips were made using a similar procedure as described in Ref.~\citenum{narasiwodeyar2015two} with reproducibly high quality and the sample surface was very smooth as shown in Sect. I.

\begin{figure}[tbp]
 \includegraphics[width=\linewidth]{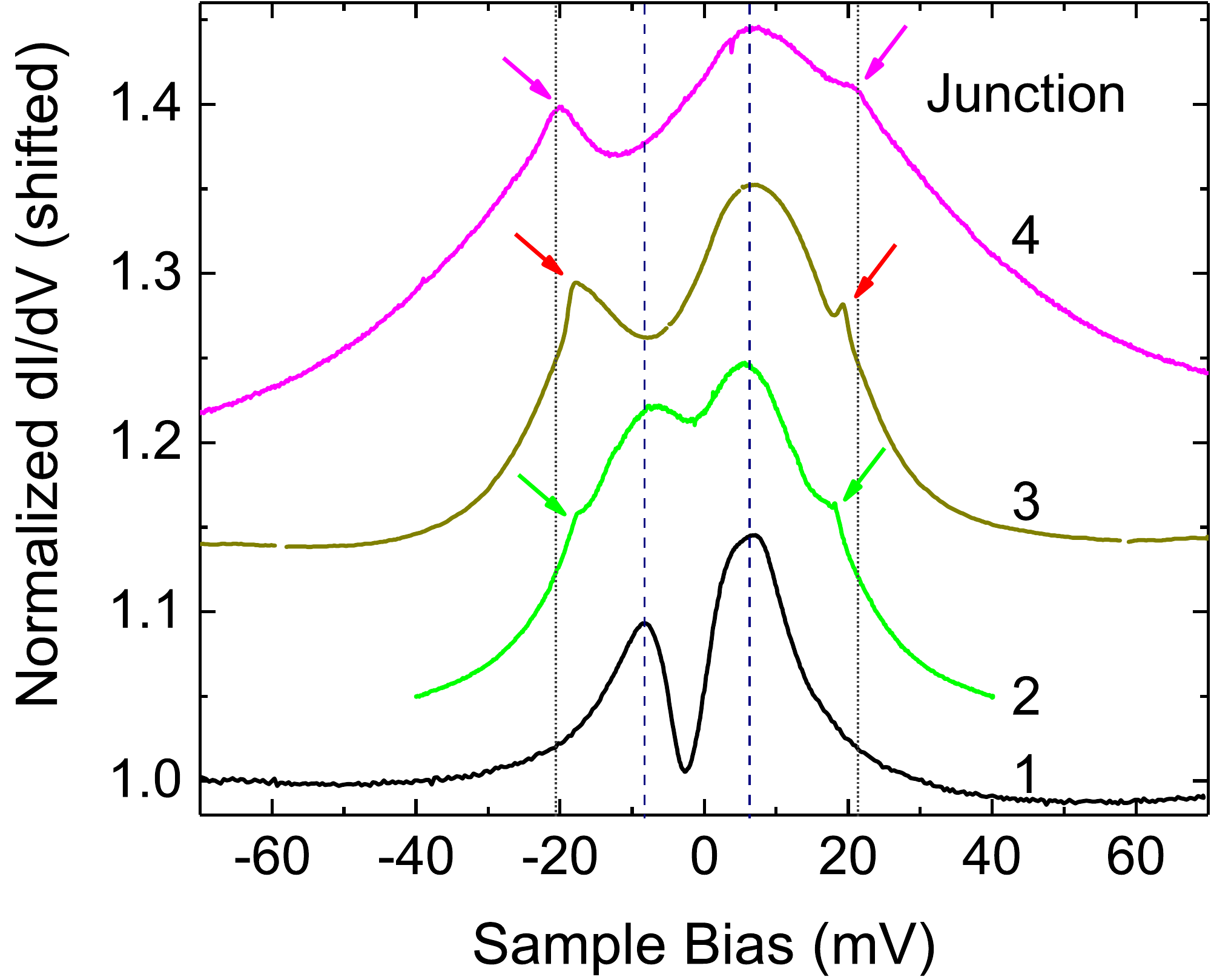}%
 \caption{Comparison of the normalized differential conductance between junctions exhibiting non-spectroscopic features (Junction 2 to 4) and a purely spectroscopic junction (Junction 1). Arrows indicate the pair of additional peaks occurring due to the local heating effect. The vertical dashed and dotted lines are a guide to the eyes for an easier comparison of the peak locations. Curves (except for Junction 1) are shifted vertically for clarity.}
 \label{fig:BallivsDiff}
\end{figure}

\begin{figure}[tbp]
 \includegraphics[width=\linewidth]{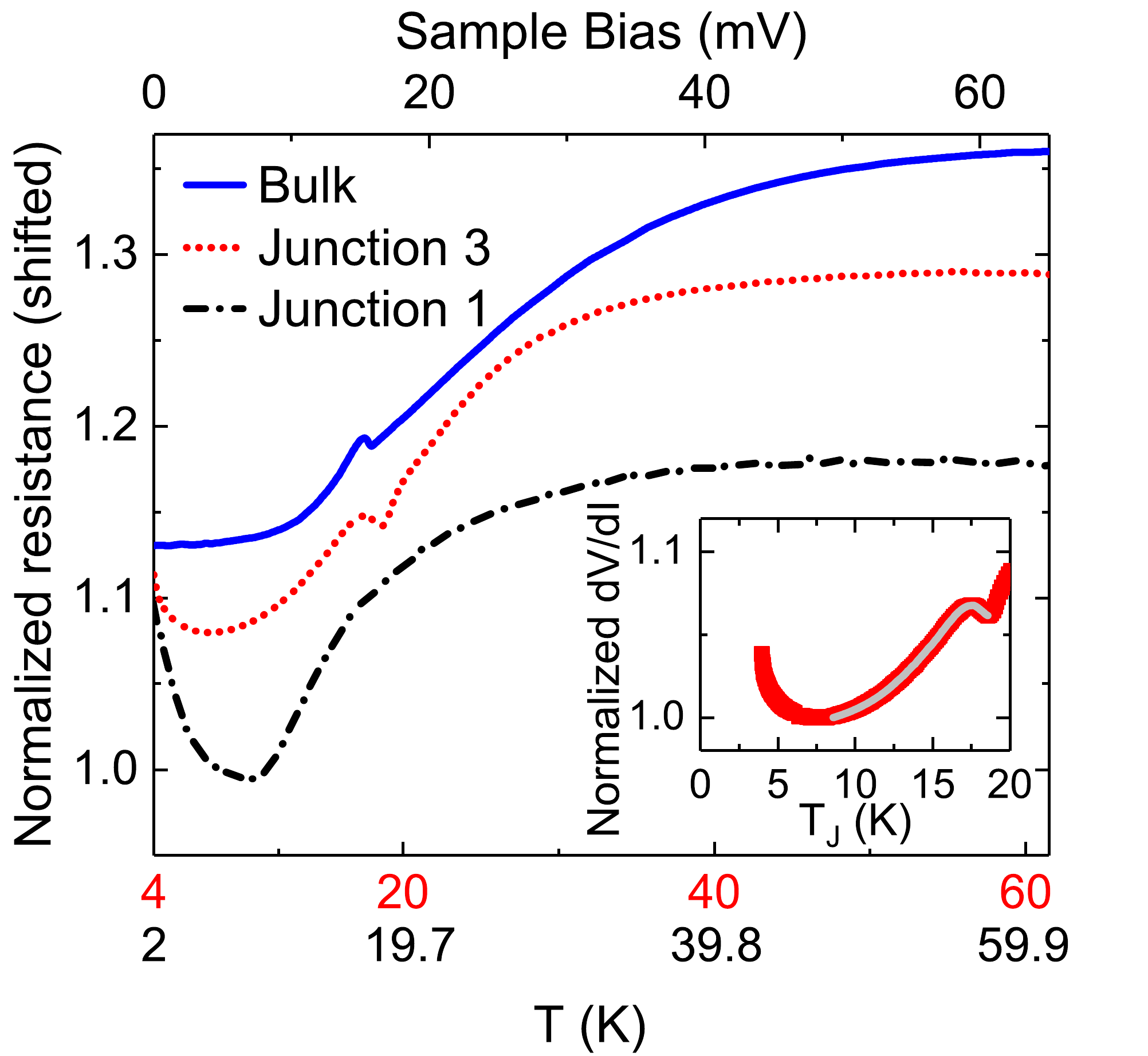}%
 \caption{Comparison of the normalized dV/dI of Junctions 1 and 3 with the bulk resistance. The bias voltage is converted to temperature via Eq.~\ref{eq:thermal} as described in the text. The upper (lower) tick labels on the bottom axis indicate the temperature for the bulk resistance and the dV/dI of Junction 3 (Junction 1). Curves are shifted vertically for clarity. Inset: Normalized dV/dI of Junction 3 as a function of junction temperature (red symbols). The gray line is a fit to Eq. 1 (from Jobiliong \etal \cite{jobiliong2005magnetization}) in the main text.}
 \label{fig:contrast}
\end{figure}

We now discuss what conductance features can be attributed to the non-spectroscopic nature of a junction on URu$_2$Si$_2$. By definition, a junction falls to the non-spectroscopic (also called thermal or Maxwell) limit if its size is much larger than the $inelastic$ mean free path, which means quasiparticles undergo inelastic scattering within the junction area. The resulting local Joule heating, whose amount increases with bias voltage, causes the junction temperature to rise as given by the following expression for the simplest case \cite{naidyuk2005point}:
\begin{gather}
T_J^2 \sim T_{bath}^2 + \frac{V^2}{4L}, \label{eq:thermal}
\end{gather}
where T$_J$ is the junction temperature, T$_{bath}$ is the bath temperature, V is the bias voltage and L is the Lorenz number. Then, the differential resistance as a function of bias voltage closely mimics the temperature-dependent bulk resistance. To illustrate this local heating effect, conductance data from four different junctions are compared in Fig.~\ref{fig:BallivsDiff}. While junction 1 shows only the aforementioned double-peak structure, the others exhibit an additional pair of little peaks (or kinks, indicated by the arrows), which has been observed to appear $always$ $outside$ the main (hybridization gap) peaks. This additional feature is of non-spectroscopic origin for the following reasons. First, the peak location varies from junction to junction, so it can't represent an intrinsic energy scale characteristic of URu$_2$Si$_2$ such as the hybridization gap, in sharp contrast to the $asymmetric$ double-peak structure whose location varies very little, as indicated by the dashed and dotted lines. Sometimes, these two types of peaks can be merged at the same bias (typically in the negative bias branch), leading to a pointy (rather than rounded) peak, as shown by the conductance curves taken from junction 3 and 4. Second, the corresponding differential resistance (dV/dI) resembles the bulk resistance, as shown in Fig.~\ref{fig:contrast}, indicative of the local heating effect. This interpretation is further supported by the fact that the two peaks are $symmetric$ in bias voltage, as expected from Eq.~\ref{eq:thermal}. Such junctions oftentimes exhibit the asymmetric double-peak structure as well (e.g., see Junction 2 in Fig.~\ref{fig:BallivsDiff}). This can be understood as due to the less severe local heating effect at low bias since practically it depends on the $local$ $thermal$ $profile$ governed by how deeply the junction falls into the thermal regime and other factors including the junction geometry.

The two different types of junctions are further compared by plotting their dV/dI vs.~V along with the temperature-dependent bulk resistance, as shown in Fig.~\ref{fig:contrast}. For a quantitative illustration of the second point mentioned above, the bias voltage is converted to temperature via Eq.~\ref{eq:thermal} using the Lorenz number L = 11.5L$_0$ \cite{behnia2005thermal}, where L$_0$ = 2.44 $\times$ 10$^{-8}$ V$^2$K$^{-2}$ is the Sommerfeld value. This Lorenz number is chosen such that the local-heating peak location matches with T$_{\rm{HO}}$ (17.5 K) after the conversion. The ticks on the temperature (bottom) axis have different labels for the two junctions because the measurement (bath) temperature was different between the two junctions: 2 K for Junction 1 and 4 K for Junction 3. As the local heating effect gets severe at high bias (see the region above 10 mV), while the dV/dI of Junction 1 increases monotonically as expected from the MDC model, the curve for Junction 3 shows a jump followed by an increase, strongly mimicking the bulk resistance. As discussed above, this additional feature is not of spectroscopic origin but due to the local heating effect. More quantitatively, the zero-temperature gap value extracted from the same analysis as in the main text (fitting to Eq.~1), as shown in the inset of Fig.~\ref{fig:contrast}, is $\Delta_0$ = 1.93 meV, smaller than the value (4.0 meV) extracted from the bulk resistance. This discrepancy may occur due to multiple factors including a more complicated thermal profile of the junction than what is assumed in deriving Eq.~\ref{eq:thermal}. Likewise, the jump-decay structure in dV/dI originating from the local heating effect can vary from junction to junction depending on the junction-specific thermal profile, as seen in Fig.~\ref{fig:BallivsDiff}. We note that none of the conductance spectra presented in the main text show this additional structure due to the local heating effect, therefore, they must contain intrinsic spectroscopic information.

Another standard check in QPS is to compare an estimated junction size with known values of mean free paths. The resistance of each junction reported in the main text is listed in Tab.~\ref{tab:S1}. The junction size is estimated by using the expressions for both Sharvin (ballistic) and Maxwell (thermal) limits, respectively: $R_J =\frac{4\rho l}{3\pi a^2}$; $R_J = \frac{\rho}{2a}$. Taking the junction on x = 0 crystal in URu$_2$Si$_{\rm{2-x}}$P$_{\rm{x}}$ as an example, the estimated junction size is about 33 \AA, much smaller than the reported elastic mean free path of 1100 \AA ~\cite{nakashima2003haas}. Therefore, this junction is well within the ballistic regime. For other junctions with finite x, estimation can't be made since mean free paths are not known yet but, from the conductance shape, one can decide that they are in the spectroscopic (at least diffusive) regime.

\begin{table}[tbp]
    \begin{center}
        \caption{The junction resistance R$_J$ taken at the maximum positive bias and the Fano parameter \text{\large\ensuremath q}$_F$ extracted from the MDC analysis of the URu$_2$Si$_{2-x}$P$_x$ and URu$_{2-y}$Fe$_y$Si$_2$ data.}
            \begin{tabular}{l c c c }
                \hline
                \hline
                Sample & $x$ or $y$ & \text{\large\ensuremath q}$_F$ & R$_J$ ($\Omega$) \\
                \hline
                URu$_2$Si$_{2-x}$P$_x$ & 0 & 8 & 37 \\
                    & 0.02 & 9.5 & 54 \\
                    & 0.03 & 12 & 91 \\
                    & 0.1 & 90 & 36 \\
                    & 0.23 & 12 & 53 \\
                    & 0.33 & 59 & 50 \\
                \hline
                URu$_{2-y}$Fe$_y$Si$_2$ & 0.02 & 15 & 52 \\
                    & 0.07 & 11.5 & 68 \\
                    & 0.17 & 14 & 83 \\
                \hline
                \hline
            \end{tabular}
            \label{tab:S1}
    \end{center}
\end{table}

\begin{figure}[tbp]
\includegraphics[width=\linewidth]{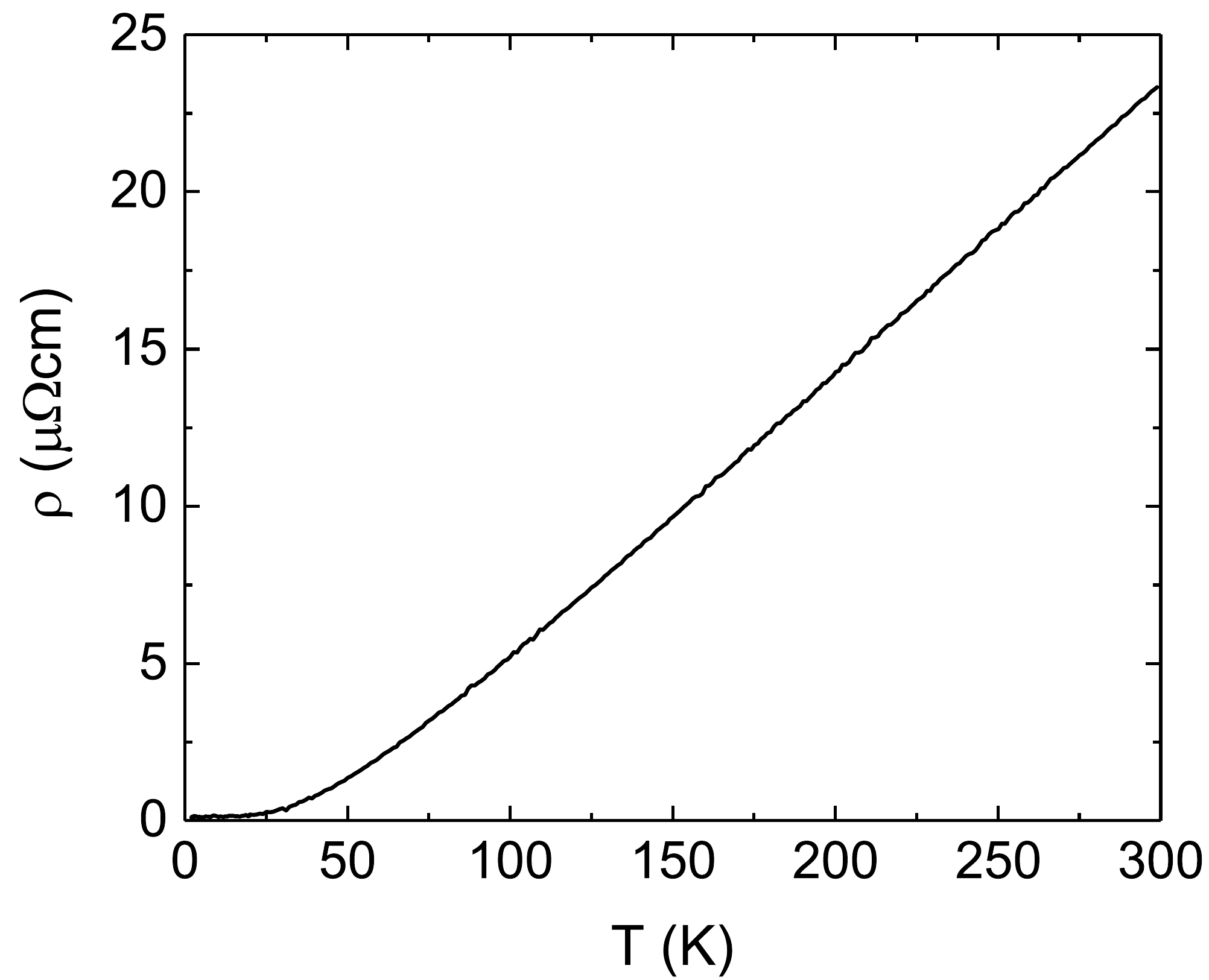}%
\caption{Temperature-dependent resistivity of ThRu$_2$Si$_2$.}
\label{fig:STh}
\end{figure}

\begin{figure*}[tbp]
\includegraphics[width=\textwidth]{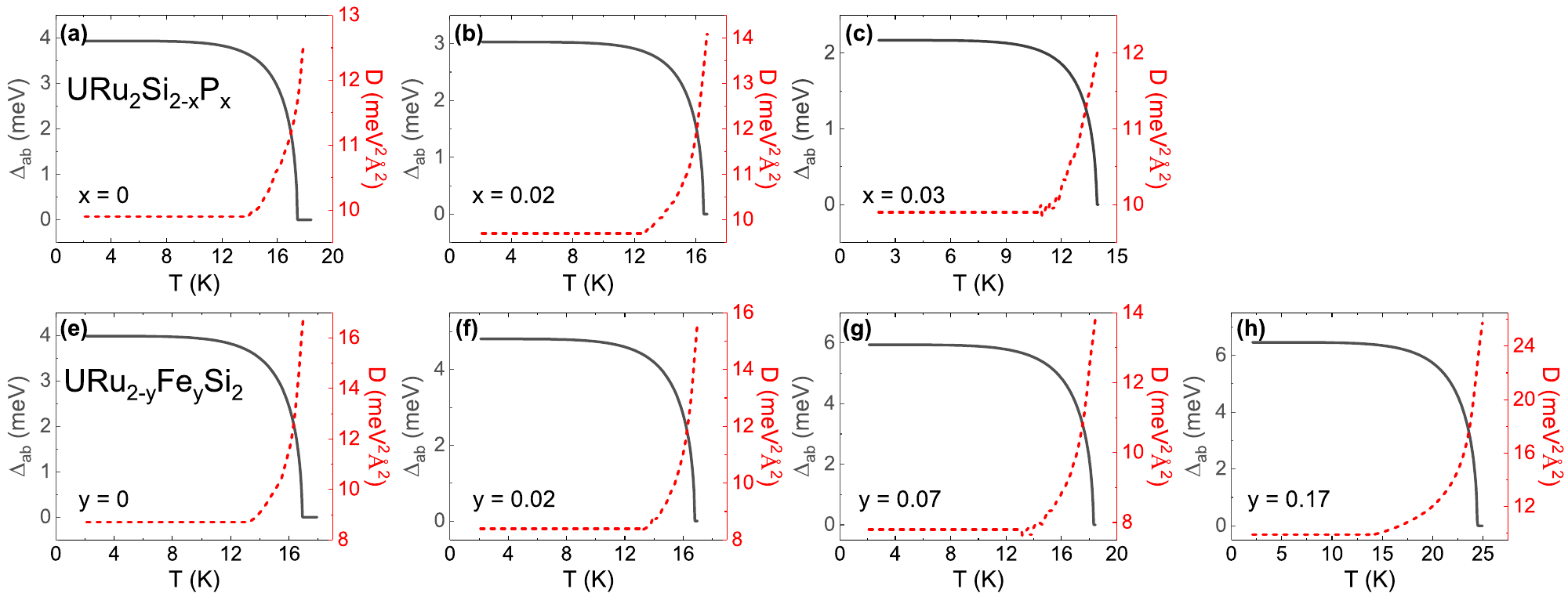}%
\caption{(a)-(d) Temperature evolution of the bosonic excitation gap $\Delta_{ab}$ (solid lines) and the stiffness $D$ (dashed lines) for URu$_2$Si$_{\rm{2-x}}$P$_{\rm{x}}$ (x = 0, 0.02, 0.03). (e)- (h) The same for URu$_{\rm{2-y}}$Fe$_{\rm{y}}$Si$_2$ (y = 0, 0.02, 0.07, 0.17).}
\label{fig:DeltaND}
\end{figure*}

\section{III. Analysis of QPS Data}
Our QPS data were analyzed using a model developed by Maltseva, Dzero and Coleman (MDC) \cite{maltseva2009electron}, in which the tunneling conductance due to electrons co-tunneling into a Kondo lattice is given as:

\begin{gather}
\frac{dI}{dV}|_{FR} \propto Im \tilde{G}^{KL}_{\Psi}(eV) \nonumber \\
\tilde{G}^{KL}_{\Psi}(\omega) = \Big(1 + \frac{\text{\large\ensuremath q}_F \Delta}{\omega - i\Gamma - \lambda}\Big)^2
\ln\Big[\frac{\omega - i\Gamma + D_1 - \frac{\mathcal{V}^2}{\omega - i\Gamma - \lambda}}{\omega - i\Gamma - D_2 - \frac{\mathcal{V}^2}{\omega - i\Gamma - \lambda}}\Big] \nonumber \\ + \frac{2 D (\frac{\tilde{t}_f}{t_c})^2}{\omega - i\Gamma - \lambda}, \label{eqn:MDC}
\end{gather}

where -D$_1$ and D$_2$ are the lower and upper edges of the conduction band, respectively. $\Delta$ is the width of Kondo resonance, $\mathcal{V}$ is the renormalized hybridization matrix amplitude, and $\tilde{t}_f$, $t_c$ are the matrix amplitudes for tunneling into $f$-level and conduction band, respectively. $\text{\large\ensuremath q}_F = \frac{\tilde{t}_f}{t_c \mathcal{V} \pi \rho}$ is the Fano parameter, governing the conductance asymmetry, as shown by relating the spectra in the main text to the corresponding $\text{\large\ensuremath q}_F$ values in Tab.~\ref{tab:S1}. The effect of $\text{\large\ensuremath q}_F$ is similar to previous QPS studies \cite{park2012observation,park2014hidden}. $\lambda$ is the renormalized $f$-level resulting from the Kondo screening process \cite{park2014hybridization}. Also, we adopted an energy-dependent broadening factor $\Gamma$ to account for the smearing in the spectra to produce better fittings \cite{wolfle2010tunneling}. The indirect hybridization gap is then determined using the following relationship: $\Delta_{hyb} = \frac{2\mathcal{V}^2}{D}$, where $2D = D_1 + D_2$ is the conduction band width.

We now discuss why the MDC model, which was formulated in the tunneling limit, can be used to analyze our QPS data. In the simplest case, differential conductance across a point-contact junction can be expressed as follows \cite{naidyuk1998point}: $\frac{dI}{dV} = \int v(E) N(E)\frac{df(E-eV)}{deV} dE$, where $v$(E) is the quasiparticle velocity and $N$(E) is the electronic density of states (DOS) in the sample. For a non-interacting system, $v$(E) and $N$(E) cancel out, resulting in Ohmic conductance, reminiscent of Harrison's theorem for the tunneling limit \cite{harrison1961}. However, it was later shown that strongly energy-dependent DOS can be detected in QPS \cite{nowack1992calculation}. This theoretical study suggests that, in a system whose DOS varies rapidly around the Fermi level, the above-mentioned cancellation no longer holds. This is particularly the case for heavy fermions since their DOS (or spectral density, more accurately) changes rapidly due to the hybridization gap. Since in this case both QPS and tunneling spectroscopy reveal the DOS, it is reasonable to consider that the same model is applicable to both cases. Furthermore, many models have been put forward to explain QPS and tunneling data taken on both single impurity Kondo and Kondo lattice systems \cite{fogelstrom2010point,fogelstrom2011tunneling,dzero2014tunneling}. In particular, the expression developed to explain the QPS conductance from a Kondo lattice heavy fermion system \cite{fogelstrom2010point} was shown to be reducible to the expression in the tunneling limit \cite{fogelstrom2011tunneling}. This can be understood qualitatively as follows: in a metallic junction on a Kondo lattice, the DOS is effectively the dominant factor since the higher order processes that contribute significantly in certain measurements such as Andreev reflection spectroscopy are negligible. For a single impurity Kondo system, it has been reported in a scanning tunneling spectroscopy study \cite{neel2007conductance} that the conductance shape evolves smoothly from tunneling to point-contact limit. In this study, the same expression for a Fano resonance \cite{fano1961u} was shown to fit to the data taken in both limits, with the only difference being the Fano parameter, $\text{\large\ensuremath q}_F$.

In any case, our main goal is to extract values for the hybridization gap and renormalized $f$-level in a systematic way. As shown in the main text and also in the previous works by some of us \cite{park2012observation,park2014hidden}, the MDC model can replicate the asymmetric double-peak structure quite nicely. Some of the fit curves in the main text do not trace the data well at high bias, but this would not affect our analysis since the high-bias features are not due to the hybridization gap.
After all, extracted parameters won't be too different regardless of the adopted model. All of this discussion indicates that our adoption of the MDC model is fairly well justified.

\section{IV. Analysis of Resistivity Data}
The phonon contribution to the resistivity of URu$_2$Si$_2$ may affect the accuracy of extracted parameters including the bosonic excitation gap because the scattering off bosonic excitations causes the same $T^5$ temperature-dependence at low temperature if the gap is zero. Thus, it is desirable to subtract out the phonon contribution first, for which resistivity instead of resistance should be used. It was too difficult to reliably define the sample dimensions since the crystals are quite small and glued onto Stycast molds for the stability during QPS measurements. Instead, assuming that the resistivity at room temperature is about the same regardless of the substituent content, we scaled the resistance with a geometric factor that gives the known resistivity at 300 K \cite{Palstra_1986} for all crystals.

Here we show in more detail how these resistivity data are analyzed. The resistivity of ThRu$_2$Si$_2$ is plotted in Fig.~\ref{fig:STh}. The RRR is about 230, indicating a high quality crystal. This resistivity is first subtracted from the URu$_2$Si$_2$ resistivity to eliminate the phonon contribution as mentioned above and in the main text. After that, the zero-temperature gap $\Delta_0$ of the bosonic excitations and the residual resistivity are extracted by fitting the low-temperature region using the Jobiliong model, Eq.~\ref{eq:approx_jobiliong} \cite{jobiliong2005magnetization}. Then, a mean-field-like temperature-dependent gap is assumed as described in the main text. Because the zero-temperature gap matches well with the E$_1$ gap in inelastic neutron scattering, $\alpha$ is set to 3.2 such that the temperature dependence of the bosonic excitation gap is similar to the E$_1$ gap. Lastly, to reproduce the resistivity at each temperature, the stiffness $D$ is set to vary freely in Eq. 1 of the main text.  $\Delta_{ab}$ and $D$ are plotted in Fig.~\ref{fig:DeltaND}. The temperature dependence of $\Delta_{ab}$ and $D$ are similar for all samples in HO and AF-II states. For the sample in AF-I state, since it's more likely a local-moment antiferromagnet as described in the main text, the resistivity is fit by the following expression \cite{jobiliong2005magnetization} with the same $\alpha$ (3.2), which is valid only in the low temperature limit ($T \ll \Delta$):

\begin{gather}
\rho(T) = \rho_0 + AT^2+ B\Delta^5 \Big[\frac{1}{5}\Big(\frac{T}{\Delta}\Big)^5 + \Big(\frac{T}{\Delta}\Big)^4 + \frac{5}{3}\Big(\frac{T}{\Delta}\Big)^3\Big] e^{-\frac{\Delta}{T}} \label{eq:approx_jobiliong}
\end{gather}

\begin{figure}[tbp]
\includegraphics[width=0.8\linewidth]{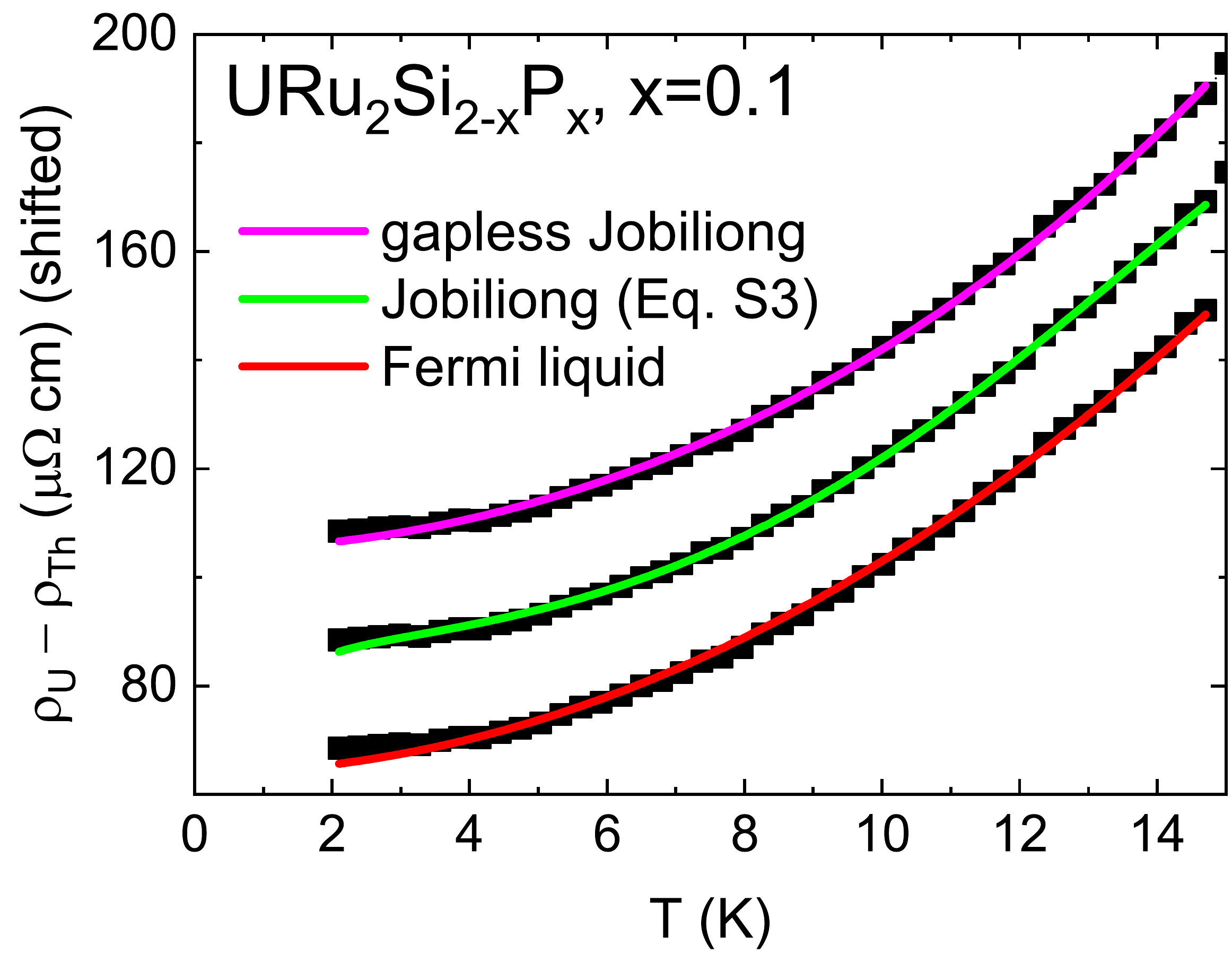}%
\caption{Fit curves to resistivity data for x = 0.1 in the NO region of URu$_2$Si$_{\rm{2-x}}$P$_{\rm{x}}$ obtained by using three expressions. Curves are shifted vertically for clarity. The similarity of fit quality indicates that bosonic excitations (other than phonons, whose contribution is subtracted out in the data) do not exist in this crystal, as expected.}
\label{fig:NOfit}
\end{figure}

For the NO region, fit curves using the following three expressions are compared in Fig.~\ref{fig:NOfit}: the Jobiliong model with constant gap, the Jobiliong model with $\Delta$ = 0, and the Fermi liquid expression, $\rho = \rho_0 + AT^2$. No apparent difference is observed in the fit quality, suggesting that the contribution from the third term is negligible. This indicates that the bosonic excitations do not exist in the NO region, as expected.

\begin{figure}[tbp]
\includegraphics[width=0.9\linewidth]{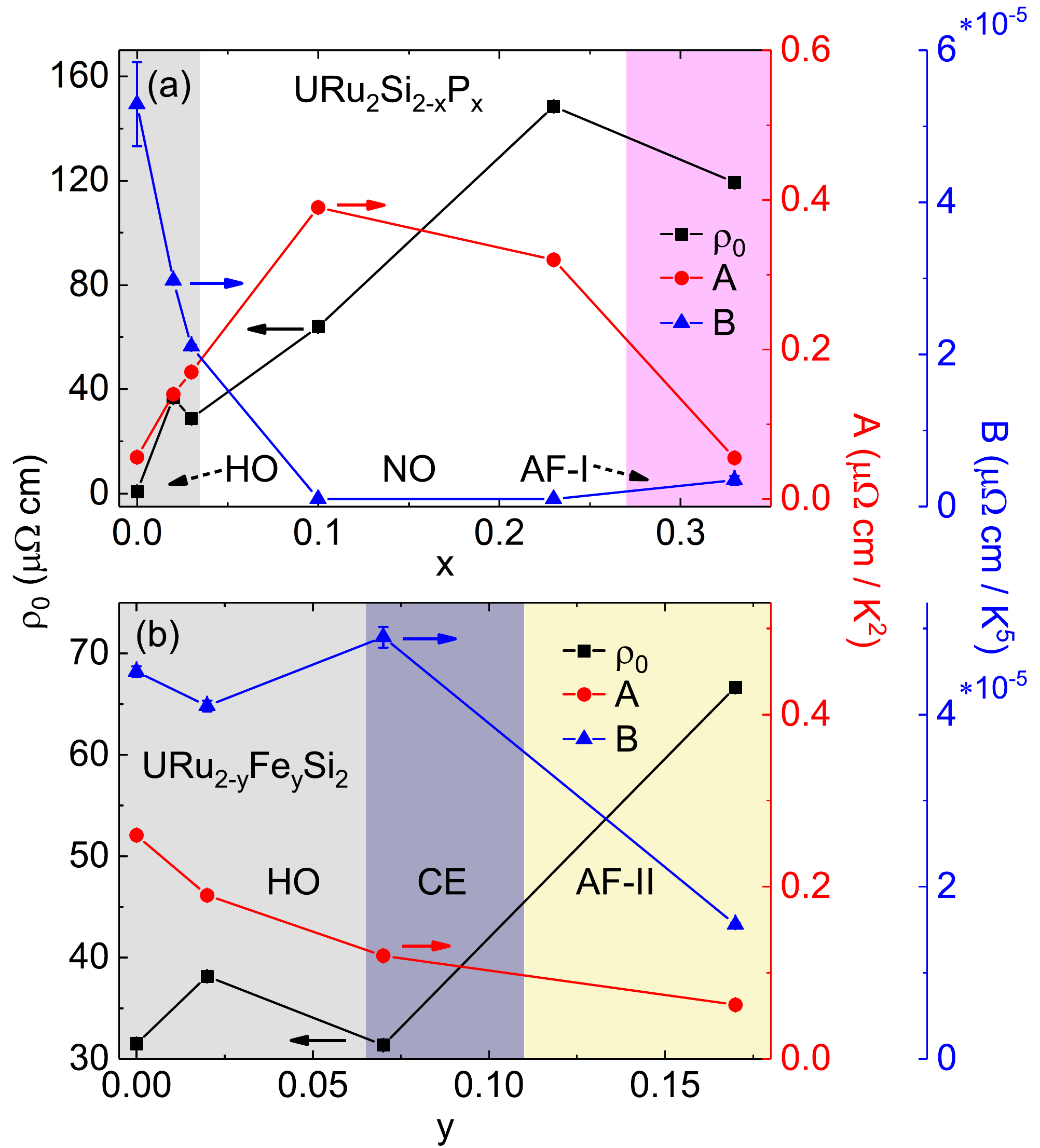}%
\caption{Substituent-content dependence of fitting parameters $\rho_0$, $A$, and $B$ extracted from our analysis of the resistivity data shown in Figs. 4(b) \& 4(e). (a) for URu$_2$Si$_{\rm{2-x}}$P$_{\rm{x}}$ and (b) for URu$_{\rm{2-y}}$Fe$_{\rm{y}}$Si$_2$.}
\label{fig:rho0_A_B}
\end{figure}

The rest of fitting parameters other than $\Delta$ and $D$, namely, $\rho_0$, $A$, and $B$ are plotted in Fig.~\ref{fig:rho0_A_B} as a function of P- or Fe-content. With increasing substituent content, $\rho_0$ increases overall, indicative of the correspondingly increasing disorder. The deviation seen for some substituent contents should be viewed cautiously since its origin may not be intrinsic due to the uncertainty in $\rho_0$ as it is not directly measured, as mentioned above, or the sample variation depending on exact growth conditions. The parameter $A$, which could be associated with the effective mass of charge carriers, exhibits discrepant behaviors between the two families of crystals. Its value for the parent compound falls in the range reported in the literature \cite{Palstra_1986,matsuda2011details,motoyama2008electrical}, 0.1 - 0.17 $\mu \Omega$cm/K$^2$. However, it shows a large variation between the two crystals (x = 0 and y = 0) used in our study and also among the literature reports, whose origin is unknown. It may be partly because several different expressions, most of which are shown in this paper, have been adopted to analyze the resistivity.

\begin{figure}[tbp]
 \includegraphics[width=0.8\linewidth]{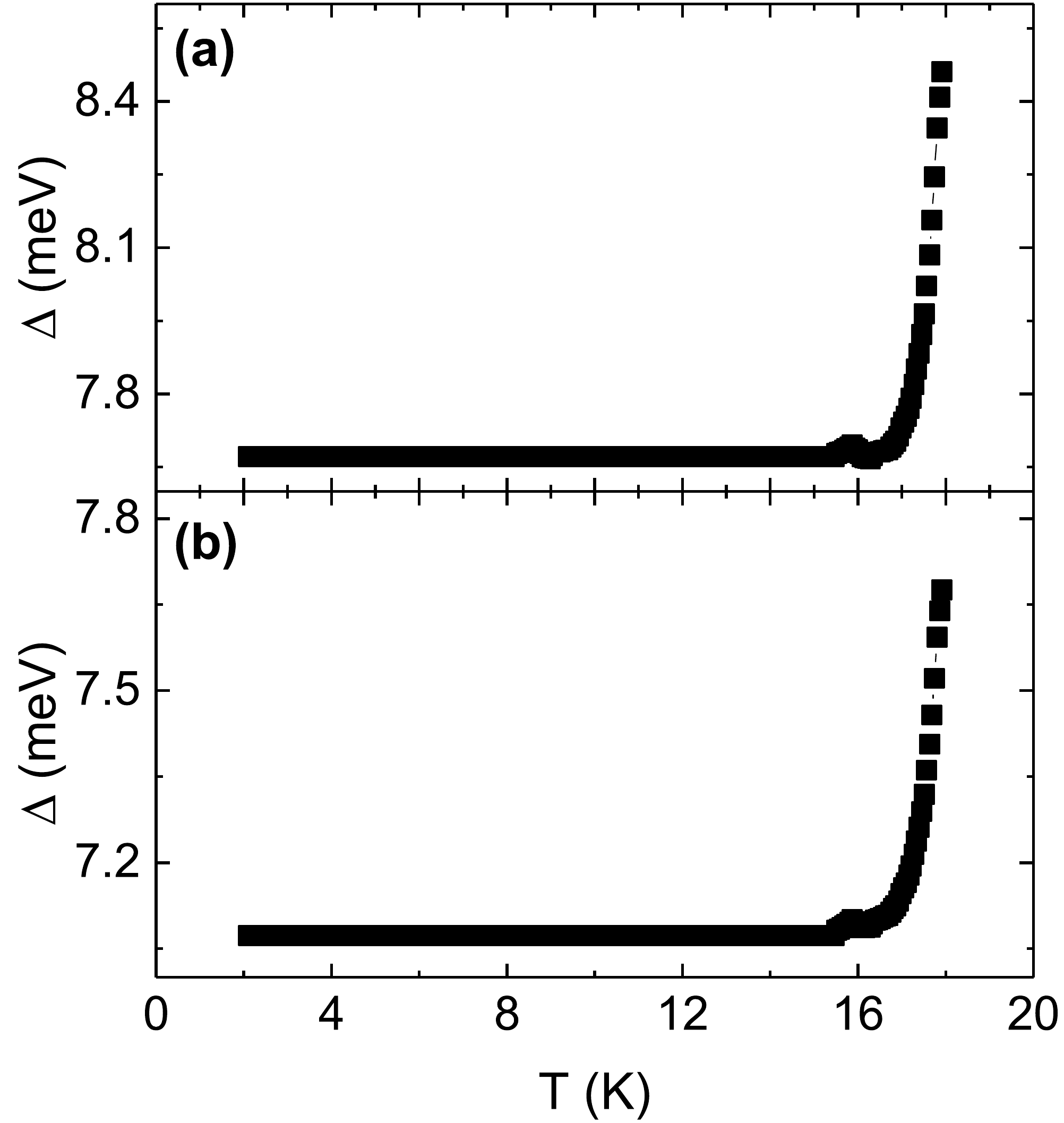}%
 \caption{(a) Bosonic excitation gap as a function of temperature extracted from a fit of the resistivity to Eq.~\ref{eq:Fontes}. (b) Same as (a) but to Eq.~\ref{eq:andersen}. The resistivity data used here are from the $x$ = 0 crystal of URu$_2$Si$_{2-x}$P$_x$.}
 \label{fig:resistivity_fit}
\end{figure}

Other expressions compared with the one used in the main text are as follows:

\begin{gather}
\rho = \rho_0 + AT^2 + B\Delta(T)^2\sqrt{\frac{T}{\Delta(T)}}\times \nonumber\\
\big[ 1 + \frac{2}{3}\big(\frac{T}{\Delta(T)}\big) + \frac{2}{15}\big(\frac{T}{\Delta(T)}\big)^2\big]e^{-\frac{\Delta(T)}{T}} \label{eq:Fontes} \\
\rho = \rho_0 + AT^2 + B\Delta(T) T\big[1+\frac{2T}{\Delta(T)}\big]e^{-\frac{\Delta(T)}{T}}, \label{eq:andersen}
\end{gather}
where Eq.~\ref{eq:Fontes} is for antiferromagnets \cite{fontes1999electron} and Eq.~\ref{eq:andersen} is for ferromagnets \cite{andersen1980crystalline}. The temperature dependence of $\Delta$ using these expressions are shown in Fig.~\ref{fig:resistivity_fit}. In order to reproduce the resistivity including the jump, both expressions should allow a rapid increase in $\Delta$ as temperature increases to T$_{\rm{HO}}$, which is not physical.

\bibliography{Supp}